\documentclass[lettersize,journal,twocolumn]{IEEEtran}
\IEEEoverridecommandlockouts
\usepackage{cite}
\usepackage{amsmath,amssymb,amsfonts}
\usepackage{graphicx}
\usepackage{subfloat}
\usepackage{subcaption}
\usepackage{multirow} 
\usepackage{makecell}
\usepackage{textcomp}
\usepackage{algorithm}
\usepackage{algorithmicx}
\usepackage{algpseudocode} 
\usepackage[T1]{fontenc}
\usepackage{xcolor}
\usepackage{amssymb}
\usepackage{booktabs}
\usepackage{color}
\usepackage{mathrsfs}
\usepackage{bm}
\usepackage{hyperref}

\usepackage{soul}
\sethlcolor{yellow}

\def\BibTeX{{\rm B\kern-.05em{\sc i\kern-.025em b}\kern-.08em
		T\kern-.1667em\lower.7ex\hbox{E}\kern-.125emX}}
\begin{document}
	
	\title{Multi-User Localization via Active Sensing with Electromagnetically Reconfigurable Antennas}
	
	\author{Ruizhi Zhang, Yuchen Zhang,~\IEEEmembership{Member,~IEEE}, Ying Zhang, Henk Wymeersch,~\IEEEmembership{Fellow,~IEEE}
		\vspace*{-0.03em}
		\thanks{R. Zhang, Y. Zhang are with School of Information and Communication Engineering, University of Electronic	Science and Technology of China, Chengdu 611731, China (e-mail: rz.uestc@gmail.com; zhying@uestc.edu.cn).
			
		Y. Zhang is with the Electrical and Computer Engineering Program, Computer, Electrical and Mathematical Sciences and Engineering (CEMSE), King Abdullah University of Science and Technology (KAUST), Thuwal 23955-6900, Kingdom of Saudi Arabia (e-mail: yuchen.zhang@kaust.edu.sa).
		
		H. Wymeersch is with the Department of Electrical Engineering, Chalmers University of Technology, Sweden (e-mail: henkw@chalmers.se).

		}
		
		
	}
	
	
	
	\maketitle
	
\begin{abstract}
	This paper investigates multi-user localization in uplink wireless systems assisted by electromagnetically reconfigurable antennas (ERAs). Unlike traditional localization schemes, we formulate an active sensing problem where a base station (BS) exploits historical pilot observations accumulated over previous sensing stages to adapt the shared ERA configuration and progressively refine position estimates. 
	To capture both theoretical flexibility and practical hardware constraints, we establish a unified wideband geometric signal model accommodating two complementary ERA paradigms: a synthesis-based model utilizing spherical-harmonic basis functions, and a finite-state model based on measured radiation codebooks. 
	Because analytically solving the resulting joint design problem is highly intractable due to the high-dimensional observation and the shared-aperture coupling among multiple users, we develop a learning-based active sensing framework. 
	Specifically, pilot-matched wideband observations are compressed into compact user-wise features and sequentially accumulated by a long short-term memory (LSTM) module. These temporal features are then processed by a graph neural network (GNN) to capture multi-user shared-aperture coupling. Model-specific output heads generate either continuous synthesis coefficients or finite-state ERA selections, while a localization head produces stage-wise position estimates. Numerical results under a specific channel distribution show that the proposed ERA-assisted active sensing framework achieves progressive localization refinement across sensing stages and obtains better performance than conventional non-reconfigurable arrays and representative ablation baselines. 
	\vspace*{-0.5em}
\end{abstract}
	
	
\section{Introduction}\label{sec:intro}


Accurate and reliable wireless localization has emerged as a key ability for next-generation 6G networks~\cite{b1,b2}. It enables a multitude of location-critical applications, including autonomous driving and robotic navigation to immersive extended reality (XR)~\cite{b3,b4}. To support these services, future localization systems must achieve ultra-high precision, even in complex propagation environments characterized by dense multi-user deployments and multipath fading~\cite{b6}.

To achieve high-precision localization, existing base stations (BSs) typically rely on massive multiple-input multiple-output (MIMO) technology~\cite{m1}. By deploying large antenna arrays, MIMO systems provide high spatial resolution and abundant spatial degrees of freedom (DoFs), enabling the separation of closely spaced users and multipath components~\cite{m2,m4}. However, conventional MIMO systems typically employ static and fixed-pattern antenna elements. The spatial flexibility of the MIMO is purely derived from adjusting the baseband or radio frequency (RF) phase shifts across the array~\cite{m5}. Consequently, the sensing capability of the MIMO is fundamentally constrained by the static electromagnetic (EM) characteristics.

To overcome the above limitations, electromagnetically reconfigurable antennas (ERAs) have recently emerged as a promising technology for next-generation wireless systems~\cite{era1,era4,era5,era6,bc-era1,bc-era2}. Unlike conventional antenna arrays with fixed electromagnetic responses, ERAs retain a fixed physical location while reshaping their metallic patterns or dielectric substrates to provide on-demand control over operating frequency, polarization, and radiation pattern~\cite{era2}. The capability introduces an additional DoF in EM domain, enabling adaptive directional response control.
As a result, ERAs can potentially improve beam adaptability and sensing performance in complex wireless systems~\cite{era-loc1}. 
Compared with spatially reconfigurable architectures such as movable antennas or fluid antennas~\cite{ma1,bc-mfa1,fa1}, ERAs can be dynamically adjusted through EM reconfiguration without altering antennas' positions. 

Motivated by the above advantages, ERAs have been investigated for wireless sensing and localization. Early studies in~\cite{era-aoa1,era-aoa2,era-aoa3} mainly focused on one-dimensional (1-D) or two-dimensional (2-D) angle-of-arrival (AoA) estimation with reconfigurable radiation patterns, demonstrating the angular sensing potential of ERAs under relatively simple narrowband assumption. 
The works in \cite{era-loc2}
further studied single-anchor indoor localization using ERAs, showing the feasibility of low-cost ERAs-assisted localization. More recently, ~\cite{era-loc1} investigated wideband localization with radiation-pattern-based ERAs and proposed hybrid baseband (BB)/EM-domain designs, verifying the benefit of ERA reconfigurability over conventional fixed-pattern arrays. Nevertheless, the existing works mainly focus on non-adaptive or single-shot sensing designs, and do not study how a shared ERA aperture should be sequentially configured for multi-user localization based on the observations accumulated over multiple sensing stages.

Active sensing provides a natural way to address the above limitation. Instead of using a fixed sensing configuration, active sensing designs the next-stage probing strategy according to the current information state obtained from previous measurements~\cite{az3,az4,az5}. From this aspect, active sensing has been applied to beam alignment, channel acquisition, and reconfigurable intelligent surfaces (RIS)-aided sensing. In particular, RIS-aided active localization has been investigated in~\cite{ar1,ar2,ar3,ar6}, where the phase configuration is sequentially adapted according to received pilots to improve localization performance. In addition, active sensing was further combined with ERA-assisted localization in~\cite{a1}, showing that the additional DoFs offered by ERAs can provide richer measurement diversity. However, these studies are mainly developed for RIS-assisted systems or single-user localization, and therefore do not address the shared-aperture sensing design.

Nevertheless, directly extending active sensing from single-user to multi-user localization is non-trivial. Although pilot-domain inter-user interference can be largely mitigated by orthogonal pilot design and pilot matching~\cite{ar3}, multiple users remain coupled through the shared ERA aperture.
In addition, the sensing strategy must extract useful information from high-dimensional historical observations and convert it into ERA configurations.
Therefore, learning-based active sensing has been shown to be effective for parameterizing such sequential sensing policies~\cite{az3,ar1,ar2,ar3,ar6,a1}. In particular, recurrent architectures have been used to summarize historical observations~\cite{az3,ar1},
which motivates the use of a long short-term memory (LSTM) module to accumulate information across sensing stages. Meanwhile, graph-based neural architectures have shown advantages in multi-user wireless system and active sensing~\cite{a6,gnn1}, since they can model interactions among users while preserving permutation-aware processing. This motivates the use of a graph neural network (GNN) to learn the shared-aperture coupling among multiple users.
These considerations motivate the learning-based active sensing framework developed in this paper.

\begin{table}[t]
	\centering
	\caption{Comparison with closely related studies.}
	\vspace{-0.5em}
	\label{tab:related_work}
	\scriptsize
	\setlength{\tabcolsep}{2.3pt}
	\renewcommand{\arraystretch}{1.12}
	\begin{tabular}{p{0.16\linewidth} p{0.36\linewidth} p{0.41\linewidth}}
		\toprule
		Work & Main scope & Difference from this work \\
		\midrule
		\cite{era-loc1} & ERA-assisted localization with hybrid/codebook design & Does not consider shared ERA configuration for multi-user scenario. \\
		\cite{a1} & ERA-assisted active sensing for localization & Focuses on single-user localization. \\
		\cite{ar1,ar2,ar3,ar6} & RIS-aided active localization with sequential configuration & Does not consider ERA-based localization design under shared aperture. \\
		\cite{a6} & GNN-based multi-user active sensing for beam tracking & Targets multi-user beam tracking rather than localization. \\
		\bottomrule
	\end{tabular}
	\vspace{-2.0em}
\end{table}

Although ERA-assisted localization and active sensing
have been separately investigated,
existing studies have not considered ERA-assisted multi-user localization with stage-wise adaptation of a shared ERA configuration. In particular, the joint design of multi-user localization, shared-aperture active sensing, and practical finite-state ERA radiation patterns remains unexplored. Motivated by this gap, this paper investigates ERA-assisted multi-user localization via learning-based active sensing.
To clarify the relationship with related studies, Table~\ref{tab:related_work} summarizes representative ERA localization and active sensing works. The main contributions of this work are summarized as follows:
\begin{itemize}
	\item We formulate an active sensing framework for ERA-assisted multi-user localization. An ERA-driven two-timescale sensing protocol is developed, where the shared ERA configuration is generated at the stage level based on accumulated pilot observations, while multiple block-wise ERA configurations are applied within each stage to provide measurement diversity. 
	To capture both theoretical flexibility and practical hardware constraints within a unified framework, we develop a signal model that incorporates two complementary ERA paradigms: a synthesis-based model built on spherical-harmonic basis functions, under which the idealized performance reference can be investigated, and a measured finite-state selection model based on a physically realizable radiation pattern library obtained in~\cite{era1} to evaluate practical performance
	\footnote[1]{Some fabrication technologies have been developed to realize ERAs, including pixel-based parasitic layouts~\cite{era4}, electronically steerable parasitic array radiator structures~\cite{era6}, and liquid-metal fluidic implementations~\cite{era1}. Without loss of generality, we adopt the implementation and modeling paradigm in~\cite{era1}, where each element's radiation pattern is reconfigured via software-controlled fluidic actuation. It is worth noting that the proposed framework is agnostic to the specific fabrication technology.
	}. 
	
	\item We propose a learning-based active sensing framework that jointly performs stage-wise localization and ERA configuration design. The framework combines an observation encoder, an LSTM-based recurrent state update, and a GNN-based multi-user interaction module. The LSTM-based module accumulates historical sensing information across stages, while the GNN-based module produces
	user representations and a shared global sensing context for shared ERA configuration. Moreover, model-specific output heads generate either continuous synthesis coefficients or finite-state ERA selections, enabling a unified design for both ERA paradigms.
	
	\item We provide numerical comparisons
	under the considered channel distribution. The results quantify the gains brought by ERA reconfigurability and active sensing. We also compare the synthesis-based model with the measured-based
	model, showing that the practically implementable finite-state ERA model can achieve performance	close to the idealized synthesis-based model. The results also demonstrate the generality of the proposed design under moderate variations from the considered channel distribution, such as different numbers of users.
	
\end{itemize}

The remainder of this paper is organized as follows. Section~\ref{sec-model} introduces the system model, and presents two ERA modeling paradigms. Section~\ref{act-sys} presents the proposed active sensing scheme for ERA-assisted multi-user localization and formulates the corresponding joint design problem. Section~\ref{pro-mec} develops a learning-based active sensing framework and describes its training procedure. Numerical results are provided in Section~\ref{res}. Finally, Section~\ref{col} concludes the paper.


In this paper, the real and the imaginary parts of a vector/matrix are denoted as $\Re(\cdot)$ and $\Im(\cdot)$, respectively. $(\cdot)^{\mathsf T}$ and $(\cdot)^{\mathsf H}$ denote the transpose and Hermitian transpose, respectively. The notation $\|\cdot\|_2$ denotes the Euclidean norm. The operator $\operatorname{vec}(\cdot)$ stacks the columns of a matrix into a vector, and $\odot$ denotes element-wise multiplication.
In addition, we use $\mathbb{R}^{m\times n}$ and $\mathbb{C}^{m\times n}$ to denote the spaces of $m\times n$ real and complex matrices, respectively. Finally, we use $\mathcal{CN}(\cdot,\cdot)$ to denote the complex Gaussian distribution.


\section{System and Signal Model}\label{sec-model}
\subsection{System Model}

\begin{figure}[t!]
	\centering 
	\includegraphics[width=0.476\textwidth]{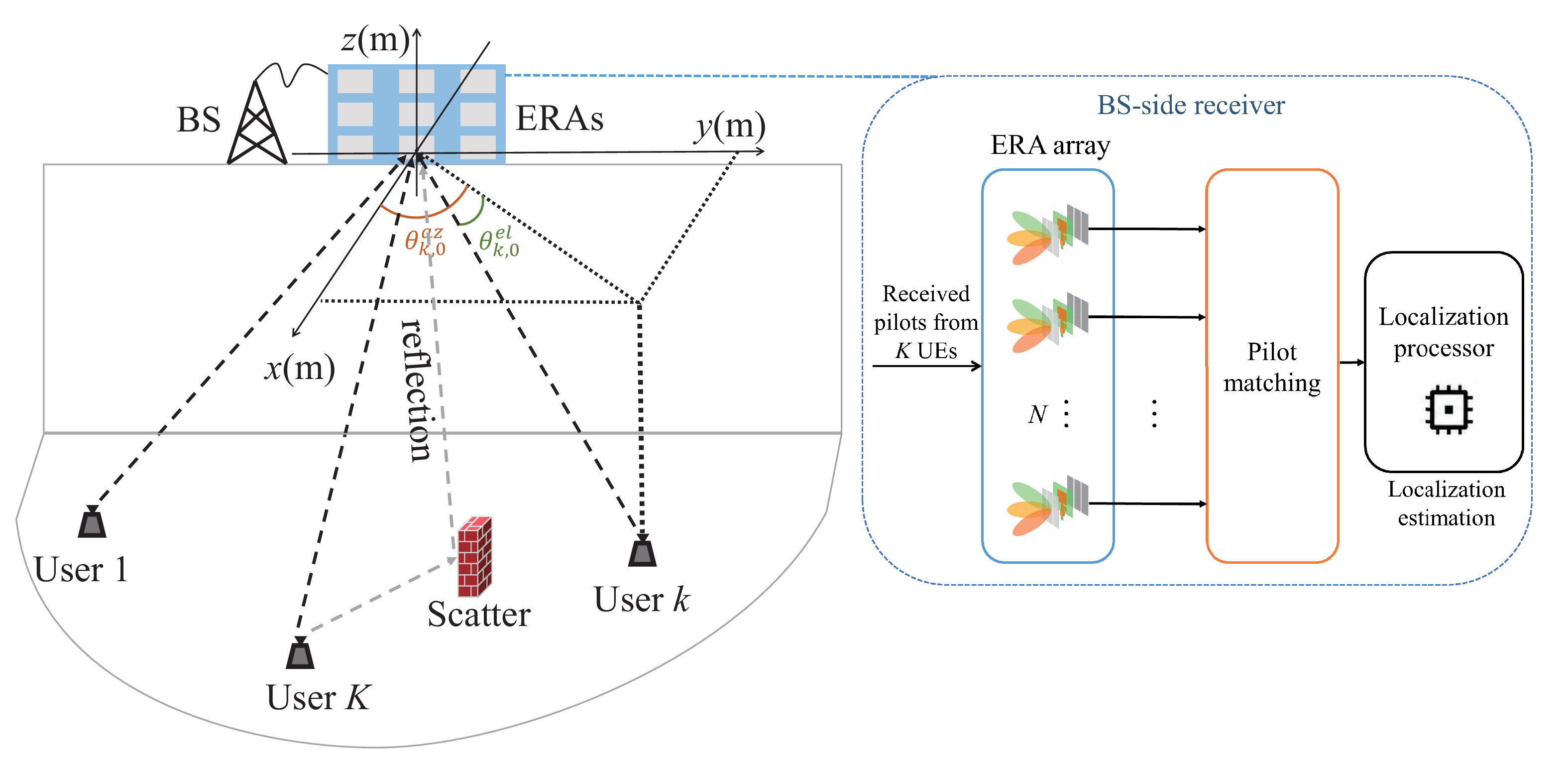} 
	\caption{Illustration of the considered ERA-assisted multi-user localization system. The ERA configuration is shared by all UEs and can be adaptively reconfigured across sensing stages.
	 }
	\vspace{-1.0em}
	\label{fig:topo}
\end{figure}

As shown in Fig.~\ref{fig:topo}, we consider a multi-user uplink localization system, where $K$ single-antenna users (UEs) with fixed positions simultaneously transmit pilot signals to a BS. The BS is equipped with $N$ ERAs to enable flexible EM reconfiguration, which provides additional spatial DoFs for the subsequent active sensing-based localization. Let $\mathbf{p}_k\in\mathbb{R}^{3}$ and $\mathbf{p}_{\mathrm B}\in\mathbb{R}^{3}$ denote the positions of the $k$-th UE and the BS, respectively. We assume that the BS and all UEs are perfectly synchronized, and synchronization errors are not considered in this work. The localization procedure is performed over $T$ active sensing stages. At each stage $t\in\{1,\ldots,T\}$, the ERA configurations are applied over $P$ pilot blocks, each consisting of $S$ pilot symbols transmitted on $M$ active subcarriers.

\begin{figure}[t!]
	\centering 
	\includegraphics[width=0.486\textwidth]{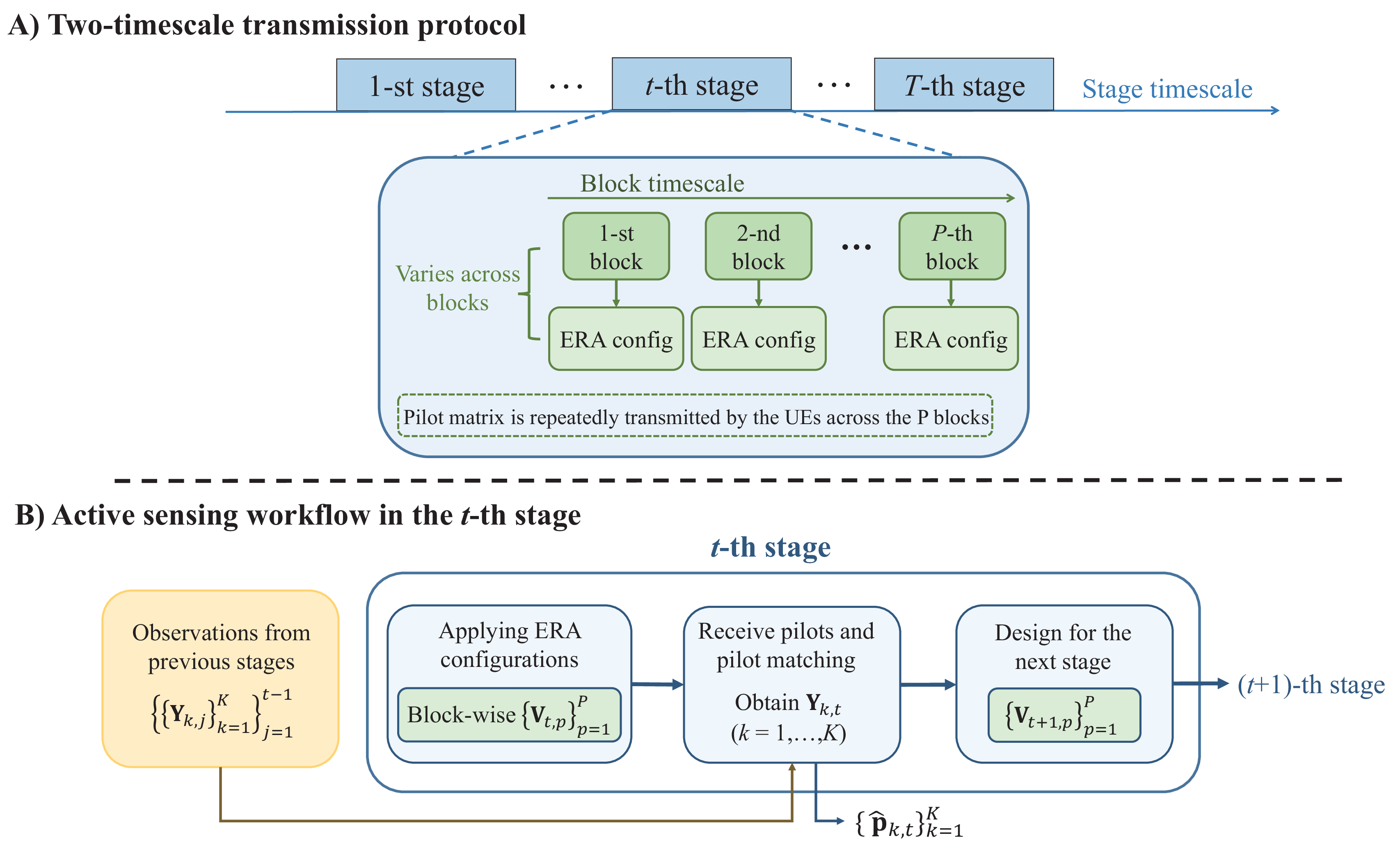} 
	\caption{
Proposed active sensing framework for ERA-assisted multi-user localization, where current-stage observations support both localization and next-stage ERA configuration.
	}
	\vspace{-1.0em}
	\label{fig:times}
\end{figure}

To enhance sensing adaptability while avoiding excessive ERA optimization and reconfiguration overhead, we adopt an ERA-driven two-timescale active sensing protocol. Specifically, the ERA configurations are designed at the stage level based on the observations accumulated from previous stages, while the generated ERA configurations are applied at the block level for the next stage. Through local EM reconfiguration, different blocks correspond to different spatial probing patterns, which provide block level measurement diversity with relatively lightweight reconfiguration overhead~\cite{era1}. In this way, the ERA-enabled BS can progressively explore the propagation environment and collect informative measurements for multi-user localization.
The proposed pilot transmission protocol is illustrated in Fig.~\ref{fig:times}. The received pilots at the $t$-th stage are used for localization and for designing the block-wise ERA configurations for the $(t+1)$-th stage.

Denote the fixed pilot matrix used in each pilot block by
\begin{equation}
	\mathbf{X} = \left[\mathbf{x}_1,\mathbf{x}_2,\cdots,\mathbf{x}_K\right]\in \mathbb{C}^{S \times K}
	\label{eq:pilot_matrix}
\end{equation}
where the $k$-th column vector 
$\mathbf{x}_k = [x_{k,1}, x_{k,2}, \ldots, x_{k,S}]^T \in \mathbb{C}^{S \times 1}$ 
contains the $S$ pilot symbols transmitted by the $k$-th UE within each pilot block.
The pilot sequences are assumed to be mutually orthogonal, i.e., 
$\mathbf{X}^{\mathrm H}\mathbf{X}=P_u S\mathbf{I}_K$, where $P_u$ denotes the uplink transmit power of each UE.
The same pilot matrix is repeatedly used over the $P$ pilot blocks in each stage.

During the $p$-th pilot block and on the $m$-th active subcarrier of the $t$-th stage, the BS first collects the antenna-domain received signal matrix
\begin{equation}
	\mathbf{R}_{t,p,m}
	=
	\sum_{k=1}^{K}
	\mathbf{h}_{k,t,p,m}\mathbf{x}_k^T
	+
	\mathbf{N}_{t,p,m},
	\label{eq:rx_signal_raw}
\end{equation}
where $\mathbf{R}_{t,p,m}\in\mathbb{C}^{N\times S}$ and
$\mathbf{N}_{t,p,m}\in\mathbb{C}^{N\times S}$ denote the received signal and additive white Gaussian noise (AWGN), respectively.
The entries of $\mathbf{N}_{t,p,m}$ are independently distributed as $\mathcal{CN}(0,\sigma^2)$ with
$\sigma^2=N_0\Delta f$, where $\Delta f$ is the subcarrier spacing.
Moreover, $\mathbf{h}_{k,t,p,m}\in\mathbb{C}^{N\times 1}$ is the effective channel from the $k$-th UE to the BS on the $m$-th subcarrier at the $p$-th pilot block of the $t$-th stage.
Since the pilot sequences are orthogonal among UEs, the BS can extract the user-specific observation for the $k$-th UE through pilot matching as
\begin{equation}
	\mathbf{y}_{k,t,p,m}
	=
	\frac{1}{P_uS}
	\mathbf{R}_{t,p,m}\mathbf{x}_k^{*}
	=
	\mathbf{h}_{k,t,p,m}
	+
	\tilde{\mathbf{n}}_{k,t,p,m},
	\label{eq:pilot_matched_signal}
\end{equation}
where $\mathbf{y}_{k,t,p,m}\in\mathbb{C}^{N\times 1}$ is the decorrelated antenna-domain observation of the $k$-th UE,
and
$\tilde{\mathbf{n}}_{k,t,p,m}
=
\frac{1}{P_uS}\mathbf{N}_{t,p,m}\mathbf{x}_k^{*}$
is the corresponding effective noise vector after pilot matching.
Since the entries of $\mathbf{N}_{t,p,m}$ are independently distributed as
$\mathcal{CN}(0,\sigma^2)$ and
$\mathbf{x}_k^{\mathrm H}\mathbf{x}_k=P_uS$, we have
$\tilde{\mathbf{n}}_{k,t,p,m}
\sim
\mathcal{CN}\left(\mathbf{0},\frac{\sigma^2}{P_uS}\mathbf{I}_N\right)$.
With orthogonal pilots, direct pilot-domain inter-user interference is removed after pilot matching.


Collecting the user-specific antenna-domain observations over all $P$ pilot
blocks, $N$ antennas, and $M$ active subcarriers within the $t$-th sensing
stage, we define the stage-wise observation matrix for the $k$-th UE as
\begin{equation}
	\mathbf{Y}_{k,t}
	=
	\left[
	\mathrm{vec}(\mathbf{Y}_{k,t,1}),
	\ldots,
	\mathrm{vec}(\mathbf{Y}_{k,t,P})
	\right]^{\mathrm T}
	\in \mathbb{C}^{P \times NM},
	\label{eq:stage_wise_observation}
\end{equation}
where $\mathbf{Y}_{k,t,p} \triangleq [\mathbf{y}_{k,t,p,1}, \mathbf{y}_{k,t,p,2},\cdots,\mathbf{y}_{k,t,p,M}]\in \mathbb{C}^{N \times M}$
denotes the block-wise observation matrix of the $k$-th UE over all $M$ active subcarriers at the $t$-th active sensing stage.
$\mathbf{Y}_{k,t}$ organizes the block level observations along its row dimension, where different rows correspond to different ERA configurations applied over different pilot blocks. Its column dimension jointly contains the antenna-domain observations and frequency-domain observations across all subcarriers.


Following the geometric model in~\cite{era1}, we represent the channel between the $k$-th UE and the BS by $L_k$ resolvable paths, then the channel vector $\mathbf{h}_{k,t,p,m}$ in~\eqref{eq:rx_signal_raw} can be given by
\begin{equation}
	\mathbf{h}_{k,t,p,m}
	=
	\sum_{\ell=0}^{L_k-1}
	\beta_{k,\ell}e^{-j2\pi\tau_{k,\ell}v_m\Delta f}
	\Big(
	\mathbf{a}(\boldsymbol{\vartheta}_{k,\ell})
	\odot
	\mathbf{g}_{t,p}(\boldsymbol{\vartheta}_{k,\ell})
	\Big),
	\label{eq:effective_channel}
\end{equation}
where $\ell=0$ corresponds to line-of-sight (LoS) component, while $\ell\in\{1,\ldots,L_k-1\}$ are non-line-of-sight (NLoS) paths generated by scatterers. $\beta_{k,\ell} = \rho_{k,\ell}e^{j\phi_{k,\ell}}$ represents the complex gain of the $\ell$-th path between the $k$-th UE and the BS, with $\rho_{k,\ell}$ and $\phi_{k,\ell}$ denoting its modulus and phase components, respectively. $\boldsymbol{\vartheta}_{k,\ell}=[\theta_{k,\ell}^{\mathrm{az}},\theta_{k,\ell}^{\mathrm{el}}]^{\mathsf T}$ denotes the azimuth-elevation direction of the $\ell$-th path observed at the BS. \(v_m \in\{-\frac{M}{2},...,\frac{M}{2}-1\}\) denote the subcarrier index.
Since the carrier frequency phase can be absorbed into the complex path gain
\(\beta_{k,\ell}\), only the baseband offset appears in the
frequency-dependent phase term. $\mathbf{a}(\boldsymbol{\vartheta}_{k,\ell})\in\mathbb{C}^{N}$ is the BS array response toward $\boldsymbol{\vartheta}_{k,\ell}$, and $\mathbf{g}_{t,p}(\boldsymbol{\vartheta}_{k,\ell})\in\mathbb{C}^{N}$ is the ERA-induced directional gain vector toward $\boldsymbol{\vartheta}_{k,\ell}$,
which will be elaborated in~\ref{sec:ERA_model} in detail. It is noted that $\mathbf{g}_{t,p}(\boldsymbol{\vartheta})$ is assumed to be frequency-flat over the considered bandwidth, since the ERA is modeled as an antenna-domain reconfiguration mechanism that changes the effective directional radiation gain of the array.
Therefore, the angle-dependent gain $\mathbf{g}_{t,p}(\boldsymbol{\vartheta})$ is evaluated at the carrier frequency and shared by all subcarriers, provided that the ERA hardware does not exhibit strong frequency selectivity over the active band.

In \eqref{eq:effective_channel}, $\tau_{k,\ell}$ denotes the delay of the $\ell$-th path between the $k$-th UE and the BS. For the LoS path, we have
\begin{equation}
	\tau_{k,0}=\frac{\|\mathbf{p}_{k}-\mathbf{p}_{\mathrm B}\|_2}{c},
	\label{eq:los_delay}
\end{equation}
where $c$ is the speed of light. And for an NLoS path associated with a scatterer at $\mathbf{p}_{k,\ell}^{\mathrm{sc}}\in\mathbb{R}^{3}$, we have
\begin{equation}
	\tau_{k,\ell}
	=
	\frac{
		\|\mathbf{p}_{k}-\mathbf{p}_{k,\ell}^{\mathrm{sc}}\|_2
		+
		\|\mathbf{p}_{k,\ell}^{\mathrm{sc}}-\mathbf{p}_{\mathrm B}\|_2
	}{c},
	\quad \ell\ge 1.
	\label{eq:nlos_delay}
\end{equation}
Under the assumption of perfect synchronization,
the delay terms in~\eqref{eq:los_delay} and~\eqref{eq:nlos_delay} only contain geometric propagation delays, without additional clock bias or timing offset~\footnote[2]{A possible clock bias would enter as \(\tau_{k,\ell}+b_k\), which is beyond the scope of this work.}.

We assume that the BS employs a uniform planar array (UPA) located on the $y$-$z$ plane, with its boresight oriented along the positive $x$-axis. Let $(y_n,z_n)$ denote the coordinates of the $n$-th antenna element with respect to the array reference point. The $n$-th entry of the steering vector $\mathbf{a}(\boldsymbol{\vartheta}_{k,\ell})$ is
\begin{equation}
	\big[\mathbf{a}(\boldsymbol{\vartheta}_{k,\ell})\big]_n
	=
	\frac{1}{\sqrt{N}}
	\exp\!\Big(
	-j\frac{2\pi}{\lambda}
	\big(
	y_n\cos\theta_{k,\ell}^{\mathrm{el}}\sin\theta_{k,\ell}^{\mathrm{az}}
	+z_n\sin\theta_{k,\ell}^{\mathrm{el}}
	\big)
	\Big),
	\label{eq:upa_response}
\end{equation}
where $\lambda$ denotes the wavelength. Here, the azimuth angle $\theta^{\mathrm{az}}$ is measured in the $x$-$y$ plane from the positive $x$-axis toward the positive $y$-axis, and the elevation angle $\theta^{\mathrm{el}}$ is measured from the $x$-$y$ plane toward the positive $z$-axis.

\vspace*{-0.86em}
\subsection{Modeling Reconfigurable Gains}\label{sec:ERA_model}
Following~\cite{bc-era1} and~\cite{era-loc1},
the reconfigurable complex amplitude response of the ERA $\mathbf{g}_{t,p}(\boldsymbol{\vartheta}_{k,\ell})$ in~\eqref{eq:effective_channel}, can be described through two
modeling paradigms. First,
an idealized ERA is assumed to realize arbitrary beampatterns on demand by projecting onto a pre-defined set of orthonormal basis functions~\cite{m4,era5}, such as spherical-harmonic (SH) basis functions~\cite{era-loc1,era5,bc1}. 
Second, each ERA switches among a finite set of discrete states, where each state corresponds to a distinct radiation response. 
A representative hardware realization of this type is reported in~\cite{era1}. In the following, both the synthesis-based model and the finite-state model are incorporated into the unified signal framework in~\eqref{eq:effective_channel}.

\subsubsection{Model I (Spherical-harmonic Synthesis Model)}

In the first model, each antenna element synthesizes its directional response from
a truncated SH basis. Let $\mathbf{b}_{\mathrm{SH}}(\boldsymbol{\vartheta})=\big[Y_1(\boldsymbol{\vartheta}),\ldots,Y_{Q}(\boldsymbol{\vartheta})\big]^{\mathsf T}\in\mathbb{C}^{Q}$ denote a vector of $Q$ orthonormal basis functions.
During the $p$-th pilot block of $t$-th stage, the radiation response of the $n$-th element of the ERA is
\begin{equation}
	\left[\mathbf g_{t,p}(\boldsymbol{\vartheta})\right]_n
	=
	\mathbf{c}_{n,t,p}^{\mathsf H}\mathbf{b}_{\mathrm{SH}}(\boldsymbol{\vartheta}),
	\label{eq:sh_gain}
\end{equation}
where $\mathbf{c}_{n,t,p}\in\mathbb{C}^Q$ is the EM coefficients that assigns
weights to the basis functions to generate the corresponding ERA gain of the $n$-th element, and subject to the per-element normalization $\|\mathbf{c}_{n,t,p}\|_2=1$~\cite{era5}. By stacking $\{\mathbf{c}_{n,t,p}\}_{n,p}$, the ERA configuration
can be represented by $\mathbf{C}_t\in\mathbb{C}^{N\times Q\times P}$.

In this paper, we adopt spherical harmonic orthogonal decomposition (SHOD) functions for basis functions~\cite{OD} due to their simplicity. Under the SHOD paradigm, any radiation pattern admits an infinite-series expansion in spherical harmonics~\cite{m4}. We can truncate these bases to the first $Q$ basis functions for convenient simulations.

\subsubsection{Model II (Measured Finite-state Model)}

In the second model, the ERA gain can be described by a calibrated finite library of realizable patterns, which provides the most faithful representation of practical hardware implementations~\cite{era5,era-loc1}. It is noted that the same state library is shared by all antenna elements, but the selected state may vary with the antenna index in a specific pilot block. Let $\mathbf{b}_{\mathrm M}(\boldsymbol{\vartheta}) = \big[b_1(\boldsymbol{\vartheta}),\ldots,b_{Q_S}(\boldsymbol{\vartheta})\big]^{\mathsf T}\in\mathbb{R}_+^{Q_S}$ denote a set of $Q_S$ candidate beampatterns, where $b_s(\boldsymbol{\vartheta})$ denotes the measured amplitude response of the $s$-th ERA state. In practice, the library is available on a discrete azimuth-elevation grid.
Accordingly, the radiation response of the $n$-th element during the $p$-th pilot block of $t$-th stage can be expressed as
\begin{equation}
	\left[\mathbf g_{t,p}^{\mathrm M}(\boldsymbol{\vartheta})\right]_n
	=
	\mathbf{s}_{n,t,p}^{\mathsf T}\mathbf{b}_{\mathrm M}(\boldsymbol{\vartheta}),
	\label{eq:measured_gain}
\end{equation}
where $\mathbf{s}_{n,t,p}\in\{0,1\}^{Q_S}$ is a one-hot state-selection vector satisfying $\|\mathbf{s}_{n,t,p}\|_0=1$. It should be noted that the measured library adopted in this work contains the directional amplitude responses of the finite ERA states. State-dependent phase variations are not included in the available pattern data, while the spatial phase progression across the array is represented by the array response $\mathbf{a}(\boldsymbol{\vartheta})$. In addition, the total power of each state is assumed to be equal to $1$ to ensure the energy conservation, i.e., $\int_{0}^{2\pi}\int_{0}^{\pi} |\mathbf{b}_{\mathrm M}(\boldsymbol{\vartheta})|^2 \sin \theta^{\rm el} d\theta^{\rm el}d\theta^{\rm az} = 1$. Similarly, by stacking $\{\mathbf{s}_{n,t,p}\}_{n,p}$, the finite-state selection of all antenna elements in the $t$-th stage can be represented by $\mathbf{S}_t\in\{0,1\}^{N\times Q_S\times P}$\footnote[3]{In our implementation, the measured library contains \(Q_S=64\) states sampled on a \(721\times361\) angular grid, with azimuth 	\([-180^\circ,180^\circ]\) and elevation \([0^\circ,180^\circ]\), both with \(0.5^\circ\) spacing. For an off-grid direction, the response of each state is obtained by bilinear interpolation over the four nearest azimuth-elevation grid points.}. Element-wise mutual coupling among elements is not explicitly modeled, and the measured state library is assumed to represent the effective response.

\section{Active Sensing For Localization}\label{act-sys}

To estimate $K$ UEs' positions, we proposes an active sensing strategy that configures the ERA in the EM domain for the $P$ blocks of each stage based on pilots received in previous stages.
Within each stage, we also design the localization scheme based on the extracted features of the received pilots. 

For the first stage, no previous observation is available. We therefore initialize $\{\mathbf V_{1,p}\}_{p=1}^{P}$ using predefined configurations. For $t\geq2$, the configurations are generated according to 
\begin{equation}
	\{\mathbf V_{t,p}\}_{p=1}^P
	=
	\mathcal{F}^{(t)}\!\left(\{\{\mathbf Y_{k,j}\}_{k=1}^K\}_{j=1}^{t-1}\right).
	\label{r-s}
\end{equation}
Here, we refer to $\mathcal{F}^{(t)}$ as the ERA configuration scheme in the $t$-th stage. $\mathbf V_{t,p}$ denotes a generic notation for the block-wise ERA configuration variable, whose specific form depends on the different ERA modeling paradigm. In particular,
\begin{equation}
	\mathbf V_{t,p}
	=
	\begin{cases}
		\mathbf C_{t,p}
		\triangleq
		\left[\mathbf c_{1,t,p},\cdots,\mathbf c_{N,t,p}\right]^{\mathsf T}
		\in \mathbb C^{N\times Q},
		& \text{Model I},\\[0.3em]
		\mathbf S_{t,p}
		\triangleq
		\left[\mathbf s_{1,t,p},\cdots,\mathbf s_{N,t,p}\right]^{\mathsf T} \in \{0,1\}^{N\times Q_S},
		& \text{Model II},
	\end{cases}
\end{equation}

Next, by performing the ERA configuration, we can obtain the observation at this $t$-th stage. Then the UEs' position can be estimated as follows
\begin{equation}
	\{\hat{\mathbf{p}}_{k,t}\}_{k=1}^K = \mathcal{G}^{(t)}\!\left(\{\{\mathbf Y_{k,j}\}_{k=1}^K\}_{j=1}^{t}\right),
	\label{p-s}
\end{equation}
where $\{\hat{\mathbf{p}}_{k,t}\}_{k=1}^K$ denotes the set of position estimates,
and $\mathcal{G}^{(t)}$ is the corresponding localization scheme. The stage-wise active sensing workflow is also illustrated in part B of Fig.~\ref{fig:times}.

Based on the above analysis, the multi-user localization problem can be formulated as the joint design of the ERA configuration schemes $\mathcal{F}^{(t)}$ and localization schemes $\mathcal{G}^{(t)}$ to minimize the cumulative stage-wise localization error, i.e.,
\begin{equation}
	\begin{aligned}
		(\mathcal{P}1):\quad
		\min_{\mathcal{F}^{(t)},\,\mathcal{G}^{(t)}}
		\quad &
		\mathbb{E}\!\left[
		\sum_{t=1}^{T}\sum_{k=1}^{K}\alpha_t
		\left\|\mathbf{p}_k-\hat{\mathbf{p}}_{k,t}\right\|_2^2
		\right] \\
		\text{s.t.}\quad
		& \eqref{r-s},\ \eqref{p-s}.
	\end{aligned}
\label{prob}
\end{equation}
The objective in~\eqref{prob} is designed for progressive localization, where a position estimate is available after each sensing stage. The stage weights $\{\alpha_t\}_{t=1}^{T}$ determine the tradeoff between intermediate-stage availability and final-stage accuracy. Equal weights, i.e., $\alpha_t = 1/T, \forall t$, promote consistently reliable estimates and are suitable for online applications with possible early termination. Larger weights on later stages instead favor final-stage accuracy and allow earlier stages to emphasize exploration. Unless otherwise stated, we use equal weights to ensure usable intermediate estimates.

However, analytically solving
$\mathcal{P}1$ is highly challenging, because it involves the joint optimization of the high-dimensional mappings $\mathcal{F}^{(t)}$ and $\mathcal{G}^{(t)}$, which are coupled across sensing and localization stages. Furthermore, in the multi-user scenario, the users are coupled through the shared ERA aperture and the unified sensing configuration, thereby making both the observation process and the active sensing design
more complicated.
In addition, as the input dimensions of these mappings increase with the number of sensing stages, deriving an analytically tractable and scalable solution becomes nearly impossible.

To address the above challenges,
we propose to utilize deep neural network as a powerful function approximator~\cite{bc2} to parameterize the above two mappings. In this way, the computational complexity of the optimization is transferred to the neural network training process~\cite{bc3}. The essence is to identify a neural network architecture capable of summarizing historical observation across different stages in designing an optimal sensing strategy for the considered system.

\vspace*{-0.56em}
\section{Learning-Based Active Sensing Framework for Multiuser Localization}\label{pro-mec}

This section develops a learning-based active sensing framework for solving problem~$\mathcal{P}1$, in which the stage-wise ERA configuration and localization policies are jointly parameterized by a neural network. At each stage, matched observations of each UE are compressed into low-dimensional user-wise features, which are then processed by an LSTM to capture cross-stage temporal information and update the corresponding hidden states. A GNN further models inter-user coupling caused by the shared ERA aperture. Based on the resulting user states, task-specific output heads generate the ERA configuration for the next stage and the position estimates for the current stage. This architecture follows the active sensing principle that current measurements should both improve immediate localization and enable more informative sensing configurations in subsequent stages.


\vspace*{-1.0em}
\subsection{Observation Encoder and Recurrent State Update}
\vspace*{-0.16em}

The recurrent module aims to summarize the information accumulated up to current stage into a fixed-dimensional state vector. Rather than directly vectorizing \(\mathbf{Y}_{k,t}\), we preserve its block, antenna, and frequency structures by reshaping its real and imaginary parts into a tensor over the block and joint antenna-frequency dimensions. The resulting tensor is then processed by a convolutional observation encoder
\begin{equation}
	\mathbf{f}_{k,t}
	=
	\phi_{\mathrm{obs}}
	\!\left(
	\Re\{\mathbf{Y}_{k,t}\},
	\Im\{\mathbf{Y}_{k,t}\}
	\right),
	\label{eq:obs_encoder}
\end{equation}
where \(\phi_{\mathrm{obs}}(\cdot)\) denotes the observation encoder. 
The convolutional layers operate on both pilot-block and frequency domain, thereby capturing joint features across the block dimension and the delay dimension. 

Next, we apply a shared-weight LSTM cell independently to summarize the evolution of the observations of each UE over stages. Equivalently, the architecture contains $K$ parallel LSTMs that share the same learnable parameters. For the $k$-th UE, the hidden state $\mathbf h_{k,t}$ and cell state $\mathbf c_{k,t}$ are updated as
\begin{equation}\label{eq:lstm_update}
	\begin{aligned}
		\mathbf{c}_{k,t} &= \mathbf{b}_{k,t} \odot \mathbf{c}_{k,t-1} + \mathbf{i}_{k,t} \odot {\rm tanh}\left(\chi_c(\mathbf{f}_{k,t}) + \psi_c(\mathbf{h}_{k,t-1})\right),\\
		\mathbf h_{k,t}&=\mathbf o_{k,t}\odot \tanh(\mathbf c_{k,t})
	\end{aligned}
\end{equation}
where $\mathbf{b}_{k,t}, \mathbf{i}_{k,t}$ and $\mathbf{o}_{k,t}$ are the activation vectors of the forget gate, input gate and output gate within the $k$-th LSTM cell, respectively. The updating rules for different gates are given as $\mathbf{m}_{k,t} = \sigma\left(\chi_m(\mathbf{f}_{k,t}) + \psi_m(\mathbf{h}_{k,t-1})\right)$, where $m\in\{b,i,o\}$, and $\sigma(\cdot)$ is the element-wise sigmoid function. For each $m\in\{b,i,o\}$, all the $\chi_m$ and $\psi_m$ are fully connected layers. The parameters of the LSTM gates are shared across all UEs, while each UE maintains its own hidden state and cell state.
In this way, the recurrent module can capture and extract important features of the received pilots, while the following modules need to handle the multi-user coupling at the current stage.

\vspace*{-0.86em}
\subsection{GNN-Based Multiuser Interaction and Joint Decision}

In multi-user localization, the ERA configuration is shared by all UEs, which creates coupling among their sensing requirements. To model this interaction while preserving permutation symmetry with respect to UE ordering, we apply a GNN to the user-wise recurrent states. The shared ERA configuration is required to be permutation invariant, whereas the per-UE position estimates should be permutation equivariant. Although several set-based architectures such as Deep Sets and Set Transformers can provide these properties~\cite{deepsets,settransformer}, we adopt a complete-graph GNN as realization because it explicitly captures pairwise UE dependencies through edge features and message aggregation~\cite{bc5}.

Based on the above analysis, after the recurrent update in \eqref{eq:lstm_update}, each UE is represented by a node feature (hidden state of LSTM cell) vector $\mathbf{h}_{k,t}$. We construct a complete interaction graph among the active UEs, where each node corresponds to one UE and each edge captures the pairwise dependency between two UEs. Let $\mathbf{u}_{k,t}^{(0)}=\mathbf{h}_{k,t}$ denote the input node embedding. At the $\ell$-th message-passing layer, the message sent from $j$-th node to $k$-th node is computed as
\begin{equation}
	\mathbf{m}_{j\rightarrow k,t}^{(\ell)}
	=
	\phi_{\mathrm{edge}}^{(\ell)}
	\!\left(
	\mathbf{e}_{j,k,t}^{(\ell-1)}
	\right),
\end{equation}
where $\phi_{\mathrm{edge}}^{(\ell)}(\cdot)$ denotes the edge-update network, and
$\mathbf{e}_{j,k,t}^{(\ell-1)}$ is the pairwise edge feature constructed from the embeddings of $j$-th and $k$-th nodes. 
Specifically, $\mathbf{e}_{j,k,t}^{(\ell-1)}$ is defined as
\begin{equation}
	\mathbf e_{j,k,t}^{(\ell-1)}
	=
	\left[
	\mathbf u_{k,t}^{(\ell-1)},
	\mathbf u_{j,t}^{(\ell-1)},
	\mathbf u_{k,t}^{(\ell-1)}-\mathbf u_{j,t}^{(\ell-1)},
	\mathbf u_{k,t}^{(\ell-1)}
	\odot
	\mathbf u_{j,t}^{(\ell-1)}
	\right].
\end{equation}
Here, the edge feature combines the two UEs embeddings with their difference and element-wise product. The individual embeddings retain UE-specific information, while the difference and product provide simple representations of relative and multiplicative interactions between the two UEs.
Then the aggregated message for $k$-th UE is obtained as
\begin{equation}
	\bar{\mathbf m}_{k,t}^{(\ell)}
	=
	\frac{1}{|\mathcal{N}_k|}
	\sum_{j\in\mathcal{N}_k}
	\mathbf{m}_{j\rightarrow k,t}^{(\ell)},
\end{equation}
where $\mathcal{N}_k=\{j | j\neq k\}$ denotes the neighbor set of $k$-th node. The node state is updated by a node-update network $\phi_{\mathrm{node}}^{(\ell)}$ as
\begin{equation}
	\mathbf{u}_{k,t}^{(\ell)}
	=
	\mathbf{u}_{k,t}^{(\ell-1)}
	+
	\phi_{\mathrm{node}}^{(\ell)}
	\!\left(
	\left[
	\mathbf{u}_{k,t}^{(\ell-1)},
	\bar{\mathbf m}_{k,t}^{(\ell)}
	\right]
	\right),
\end{equation}
After $L_{\mathrm G}$ graph layers, we obtain the interaction-aware node embedding $\mathbf{u}_{k,t}^{L_{\mathrm G}}$ for each UE. For notational simplicity, we omit the superscript $L_{\mathrm G}$ when no ambiguity arises.
Based on these node embeddings, we employ two specific feed-forward heads to generate the current-stage position estimates and the ERA configurations for the next stage. Specifically,
\begin{equation}
	\begin{aligned}
		\{\hat{\mathbf p}_{k,t}\}_{k=1}^K&=\phi_{\mathrm p}(\mathbf{u}_{k,t}), \\
		\{\mathbf{v}_{t+1,p}\}_{p=1}^P&=
		\mathcal{N}_{\mathrm v}\!\big(\phi_{\mathrm v}(\mathbf{u}^{\mathrm g}_{t})\big),
	\end{aligned}
\end{equation}
where $\phi_{\mathrm p}(\cdot)$ and $\phi_{\mathrm v}(\cdot)$ denote the localization and ERA configuration heads, respectively, while $\mathbf{u}^{\mathrm g}_{t}$ is the global context vector for generating the shared ERA configuration. The normalization layer $\mathcal{N}_{\mathrm v}(\cdot)$ enforces the corresponding constraints on the EM coefficients. The configuration head is model specific: it outputs continuous ERA synthesis coefficients for Model~I and discrete codebook state-selection variables for Model~II. These output designs are detailed in subsection~\ref{ERA_output}.

\vspace*{-0.36em}
\subsection{Model-Specific ERA Output Heads}\label{ERA_output}
The above observation encoder, LSTM recurrent update module, and GNN interaction module can be shared by Model~I and Model~II under the same aperture architecture and propagation environment. The only ERA model-dependent module lies in the output parameterization of the ERA configuration head.
Since the ERA configuration is shared across UEs at the BS side, we first aggregate the user-wise graph embeddings into a global context vector
\begin{equation}
	\mathbf{u}^{\mathrm g}_{t}
	=
	\sum_{k=1}^{K}\omega_{k,t}\mathbf{u}_{k,t},
\end{equation}
where the attention weights are obtained by a shared scoring network $a_{k,t}=\phi_{\rm att}(\mathbf u_{k,t})$, where \(\phi_{\rm att}(\cdot)\) is a lightweight network.
Then the weight is obtained by $\omega_{k,t}=\frac{\exp(a_{k,t})}{\sum_{j=1}^{K}\exp(a_{j,t})}$.

\subsubsection{Model I: Continuous SH-Coefficient Head}

For the synthesis-based ERA model, the output head generates continuous SH coefficients for all antennas and pilot blocks as
\begin{equation}
	\widetilde{\mathbf C}_{t+1}
	=
	\phi_{\mathrm C}\!\left(\mathbf{u}^{\mathrm g}_{t}\right),
\end{equation}
where $\widetilde{\mathbf C}_{t+1}\in\mathbb{C}^{N\times Q\times P}$. To satisfy the per-element normalization constraint in \eqref{eq:sh_gain}, the output is normalized as
\begin{equation}
	\mathbf c_{n,t+1,p}
	=
	\frac{\widetilde{\mathbf c}_{n,t+1,p}}
	{\|\widetilde{\mathbf c}_{n,t+1,p}\|_2},
	\quad \forall n,p.
\end{equation}
The resulting $\mathbf C_{t+1}$ is then used to synthesize the directional response for the next stage.

\subsubsection{Model II: Discrete State-Selection Head}

For the measured finite-state ERA model, the output head instead generates the logits associated over candidate states
\begin{equation}
	\mathbf{\Pi}_{t+1}
	=
	\phi_{\mathrm S}\!\left(\mathbf{u}^{\mathrm g}_{t}\right),
\end{equation}
where $\mathbf{\Pi}_{t+1}\in\mathbb{R}^{N\times P\times Q_S}$ contains the state-selection scores for all antenna elements and pilot blocks. Its $(n,p,:)$-th slice, denoted by $\boldsymbol{\pi}_{n,t+1,p}\in\mathbb{R}^{Q_S}$, represents the logits over the \(Q_S\) candidate states for \(n\)-th antenna in \(p\)-th pilot block of $(t+1)$-th stage.
Because discrete state selection is non-differentiable, we employ a straight-through estimator (STE) during training to enable gradient-based optimization~\cite{ste}. Specifically, a soft state-selection vector is first constructed from \(\boldsymbol{\pi}_{n,t+1,p}\) using the Gumbel--Softmax relaxation~\cite{gs}
\begin{equation}
	\bar{\mathbf s}_{n,t+1,p}
	=
	\mathrm{GS}\!\big(\boldsymbol{\pi}_{n,t+1,p};\tau\big),
\end{equation}
where $\operatorname{GS}(\boldsymbol{\pi};\tau)$ denotes the
Gumbel--Softmax relaxation with temperature $\tau$. A hard one-hot state is then generated as
\begin{equation}
	\hat{\mathbf s}_{n,t+1,p}
	=
	\mathrm{onehot}\!\left(
	\arg\max_{s}[\bar{\mathbf s}_{n,t+1,p}]_s
	\right).
\end{equation}
Here, $\operatorname{onehot}(\cdot)$ converts an index into a one-hot vector. To preserve end-to-end trainability, the effective training state vector is formed via the straight-through~\cite{ste}
\begin{equation}
	\widetilde{\mathbf s}_{n,t+1,p}
	=
	{\rm detach}\left(\hat{\mathbf s}_{n,t+1,p}
	-
	\!\bar{\mathbf s}_{n,t+1,p}\right)
	+
	\bar{\mathbf s}_{n,t+1,p},
\end{equation}
where ${\rm detach}(\cdot)$ denotes the stop-gradient manipulation. The forward process uses the hard-selection results, while the back-propagation reuses the gradient of the soft relaxed vector. 

\subsection{Extension to Variable Numbers of UEs}
\begin{figure}[!t]
	\centering 
	\includegraphics[width=0.439\textwidth]{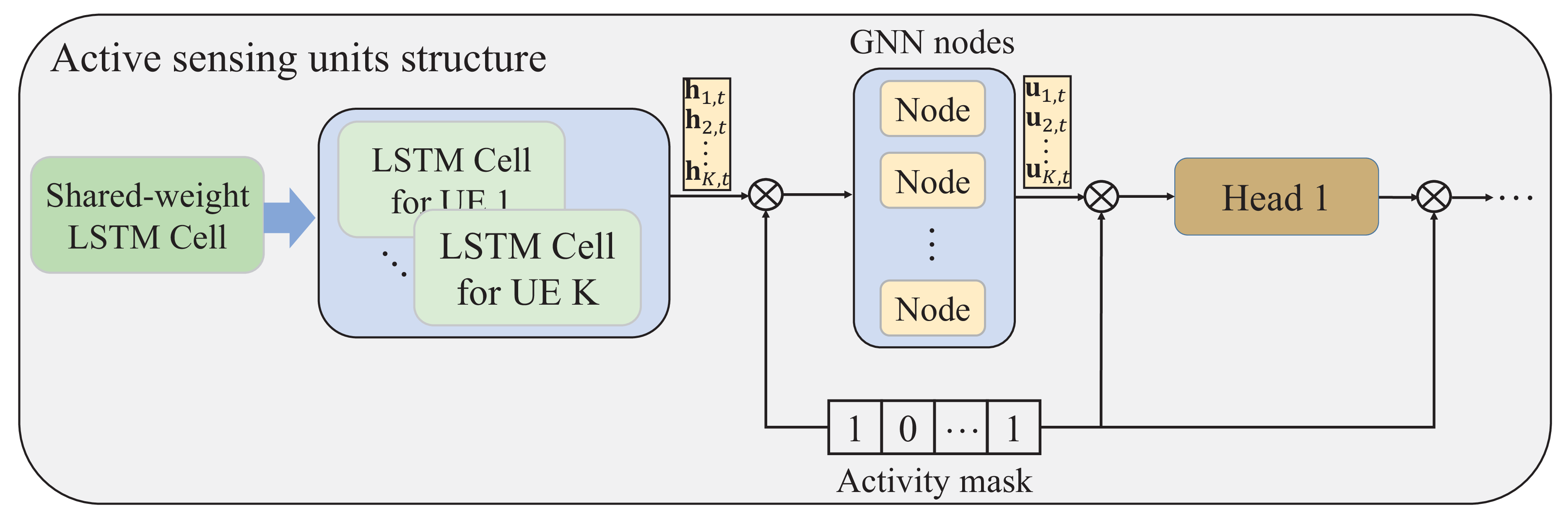} 
	\caption{Illustration of the unified active sensing units.}
	\vspace{-1.0em}
	\label{fig:share_w}
\end{figure}

\begin{figure*}[!t]
	\vspace{-0.5em}
	\centering 
	\includegraphics[width=0.86\textwidth]{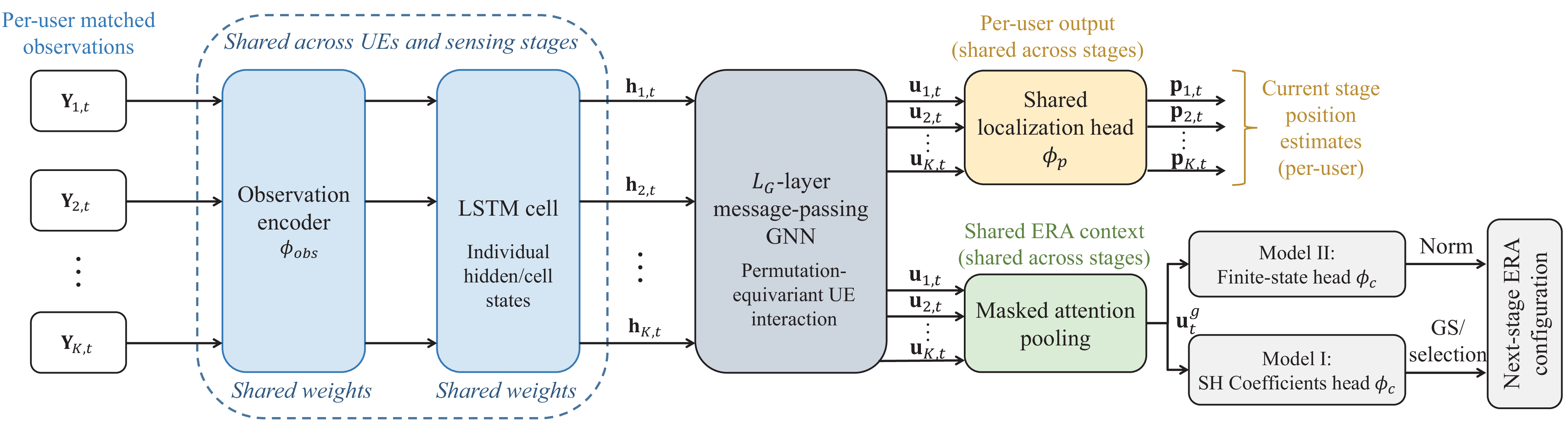} 
	\caption{Detailed architecture of the proposed learning-based model. The encoded observations are accumulated by LSTM cells and then processed by a GNN-based module to capture multi-user coupling caused by the shared ERA configuration. The output heads jointly generate the current-stage position estimate and the model-specific next-stage ERA configuration.}
	\label{fig:model}
\end{figure*}

In practical wireless systems, the number of UEs may vary with the load and service demand. Nevertheless, the proposed framework is inherently compatible with such variations, providing the possibility to achieve such generalizability.
The observation encoder and recurrent module operate independently on each UE, allowing their parameters to be shared across different UEs. Similarly, the GNN uses shared edge- and node-update networks to capture inter-user coupling induced by the shared ERA aperture and sensing configuration. The main scalability challenge therefore lies in the two output heads, as directly concatenating user features would result in input dimensions that depend on the number of UEs.


To distinguish the maximum network capacity from the actual number of active UEs, we denote $K_{\rm max}$ as the maximum number of UEs supported by the network architecture and pilot design. An activity indicator $d_{k}$ is introduced, where $d_{k} = 1$ indicates that the $k$-th UE is active. Accordingly, the number of active UEs in the sample is $K_{\rm act} = \sum_{k=1}^{K_{\rm max}}d_k$.

As illustrated in Fig.~\ref{fig:share_w}, the proposed active sensing unit is designed to accommodate different numbers of UEs within a unified architecture. Specifically, the user-wise observation encoder and LSTM cell adopt a shared-weight design, such that the same feature extraction module can be reused for all UEs. In this way, pilot observations from different UEs are processed in a consistent manner, and the feature extraction stage does not depend on the number or ordering of active UEs.
To further support variable-size multi-user scenarios, the extracted user features are organized as graph nodes and processed by a GNN, which is particularly well suited to this setting.
Under a prescribed $K_{\rm max}$, an activity mask $\mathbf{d}=[d_1,...,d_{K_{\rm max}}]^T$ is introduced to indicate which UEs are active in the current sample. The mask is used to suppress inactive nodes and restrict graph interaction, feature aggregation, masked softmax for user weighting, and output evaluation to the active UEs only. 
As a result, the same GNN and output heads can be directly applied to scenarios with different numbers of UEs without architectural changes. 

\vspace*{-0.66em}
\subsection{Training Procedure}

The active sensing units are concatenated to form a deep recurrent architecture, as illustrated in Fig.~\ref{fig:model}, where each unit corresponds to one active sensing stage. The sensing unit is recursively unrolled over $T$ stages to form an end-to-end recurrent architecture. At each stage, the current pilot observations are encoded and combined with the user-wise recurrent states, after which the GNN produces the current position estimates and the ERA configuration for the next stage. Since the same sensing policy is repeatedly applied throughout this sequential procedure, the proposed architecture adopts a structured parameter-sharing design.

Specifically, the observation encoder $\phi_{\mathrm{obs}}$ and the LSTM parameters are shared across all UEs and all sensing stages. Each UE maintains its own recurrent hidden and cell states. At the $\ell$-th 	graph layer, the edge-update network $\phi_{\mathrm{edge}}^{(\ell)}$ is shared across all directed edges, and the node-update network $\phi_{\mathrm{node}}^{(\ell)}$ is shared across all UE nodes. These networks are also reused across sensing stages, whereas different graph layers employ distinct parameters. The attention and localization heads are also shared across UEs and stages, whereas the ERA configuration head is model specific. Consequently, all sensing stages reuse the same unit, and the stage dependence is represented through the recurrent states and the accumulated observations rather than through stage-specific network. 

For variable-user training, we employ a mask-normalized implementation of the objective of $\mathcal{P}1$. Specifically, the loss for each training sample is computed as
\begin{equation}
	\mathcal{L} = \sum_{t=1}^{T} \alpha_t \frac{1}{K_{\rm act}} \sum_{k}^{K_{\rm max}} d_k\left\|\mathbf{p}_k-\hat{\mathbf{p}}_{k,t}\right\|_2^2.
\end{equation}
Therefore, inactive UEs do not contribute to the loss, and the localization error of each sample are normalized by its number of active UEs. 
The proposed architecture is trained offline in an end-to-end supervised manner using Adam to minimize the mask-normalized empirical objective of \(\mathcal{P}1\). By unrolling a sufficiently large number of stages during training, the network jointly learns the ERA configuration policy $\mathcal{F}^{(t)}$ and the localization scheme $\mathcal{G}^{(t)}$ to progressively refine sensing and position estimates from successive pilot observations.
Since the recurrent sensing unit shares parameters across stages, the trained model can be recursively applied for any number of stages up to the training length and terminated early according to the available sensing budget.


\section{Numerical Results}\label{res}

In this section, we evaluate the localization performance of the proposed framework 
through numerical simulations.

\vspace*{-1.0em}
\subsection{Simulation Scenario and Channel Generation}

The considered scenario consists of a BS equipped with a square UPA, i.e., the UPA has the same number of rows and columns.
The UEs are independently distributed within the front-side semicircular annulus of the BS.
More specifically, the UE azimuth is uniformly sampled from the front-side sector $[-\pi/2,\pi/2]$ with respect to the BS boresight, the horizontal location is sampled over the annulus with the distance range in Table~\ref{tab:simulation_parameters}, and the UE height is uniformly sampled within the specified $z$-coordinate range.
Unless otherwise specified, the default system and scenario parameters are summarized in Table~\ref{tab:simulation_parameters}, where the main settings are selected with reference to representative studies~\cite{era-loc1,a1,ar1,a6}.

\begin{table}[t]
	\centering
	\caption{System parameters}
	\label{tab:simulation_parameters}
	\begin{tabular}{|c|c|}
		\hline
		Default System Parameters & Value \\
		\hline
		BS position $\mathbf p_B$ & $[0,0,0]^T$ m \\
		UPA size & $4 \times 4$ \\
		Carrier frequency $f_c$ & 3.5 GHz \\
		System bandwidth $B$ & 100 MHz \\
		Number of active subcarriers $M$ & 256 \\
		Noise PSD $N_0$ & $-154$ dBm/Hz \\
		Number of SHOD bases $Q$ & 25 \\
		UE horizontal distance range & $[10,50]$ m \\
		UE $z$-coordinate range & $[-10,10]$ m \\
		Scatterer horizontal distance range & $[5,60]$ m \\
		Scatterer $z$-coordinate range & $[-10,10]$ m \\
		\hline
	\end{tabular}
	\vspace*{-1.05em}
\end{table}

To make the statistical channel model explicit, all training and testing samples are generated independently from the same geometric wideband channel distribution described in Section~\ref{sec-model}.
The BS and all UEs are assumed to be perfectly synchronized, and thus the path delays only include geometric propagation delays.
For each UE, the LoS component is always present, and the number of NLoS paths is fixed as $3$.
Accordingly, each UE-BS uplink channel contains $L_k=4$ resolvable paths in total.
Each NLoS path is generated by an independent single-bounce scatterer $\mathbf{p}_{k,\ell}^{\mathrm{sc}}$, whose azimuth is uniformly sampled from $[-\pi/2,\pi/2]$ and whose horizontal distance and height are independently sampled according to the scatterer ranges in Table~\ref{tab:simulation_parameters}.
Given the UE and scatterer locations, the angle and delay of each path are calculated according to the geometric relations in~\eqref{eq:los_delay} and~\eqref{eq:nlos_delay}. The complex path coefficients are generated according to a free-space propagation model.
Specifically, the LoS path amplitude is modeled as $\rho_{k,0}=\frac{\lambda}{4\pi\|\mathbf{p}_{\mathrm B}-\mathbf{p}_{k}\|}$, while the amplitude of the $\ell$-th NLoS path is modeled as
$\rho_{k,\ell}=\frac{\sqrt{4\pi}\lambda}{16\pi^2\|\mathbf{p}_{\mathrm B}-\mathbf{p}_{k,\ell}^{\mathrm{sc}}\|\|\mathbf{p}_{k,\ell}^{\mathrm{sc}}-\mathbf{p}_{k}\|}$, $\ell=1,\ldots,L_{\mathrm{NLoS}}$, following~\cite{era-loc1}.
The path phase components are independently generated from a uniform distribution over $[-\pi,\pi]$.
The ERA response is evaluated at the carrier frequency and is assumed to be frequency-flat.

\vspace*{-1.0em}
\subsection{Network Implementation and Training Setup}

During training, the number of active users $K_{\rm act}$ in each sample is randomly drawn from $3$ to $8$, and the proposed active sensing framework is trained with $T=6$ sensing stages.
Each sensing stage contains $P=4$ pilot blocks, and each block contains $10$ mutually orthogonal pilot symbols. For the first sensing stage, the $4$ blocks use predefined structured ERA probing directions. The target azimuth angles are uniformly spaced over the front sector $[-60^\circ, 60^\circ]$, with elevation fixed at $0^\circ$. For Model I, the initial probing patterns are synthesized using normalized low-order SH coefficients, where only orders $\ell \leq 1$ are retained and the coefficients of order $\ell$ are tapered by $0.35^{\ell}$. For Model II, the state of each block is selected from the measured codebook as the state with the largest gain toward the corresponding target direction. Within each block, the same initial configuration is applied to all antenna elements.

After training, since the sensing unit is shared across stages, the learned model can be recursively reused and early-terminated to support any number of stages satisfying $T_{\mathrm{test}}\leq 6$ under the fixed blocks, which allows us to evaluate its early-termination capability.
In addition, because each block contains $10$ mutually orthogonal pilot symbols, the maximum number of active UEs supported during testing is $K_{\rm max} = 10$.
Therefore, the generalization should be interpreted within the considered synchronized geometric channel distribution, rather than as robustness to arbitrary propagation environments.

For the network, the observation encoder is implemented as a lightweight 2D CNN with $36$ base channels, while the multi-user interaction module is realized by a two-layer message-passing GNN. Inter-stage information is propagated through an LSTM cell. All output heads are implemented using fully connected networks. In the implementation, the hidden feature dimension of all modules is set to $256$. The model is trained with a batch size of 128 for 50,000 iterations. To enable the stable training, the network is optimized using Adam with an initial learning rate of $10^{-3}$ and the weight decay coefficient is set to $10^{-6}$. Gradient clipping is further applied with a clipping threshold of 1.0 to improve stability~\cite{dl}. In Model~II experiments, the temperature is initialized as \(\tau_0=1.25\) and annealed to \(\tau_{\min}=0.35\). Denote \(i\) as training iteration and \(I\) the total number of iterations. We use an annealing schedule $\tau(i) = \tau_{\min}+\frac{1}{2}(\tau_0-\tau_{\min})
\left(1+\cos\left(\pi \frac{i}{0.9I}\right)\right)$.
During training, the forward pass uses hard selections through
the STE, while gradients are propagated through the relaxed
Gumbel-Softmax sample. During inference, the selected state is obtained
by the deterministic argmax over the logits. We implement the proposed framework using PyTorch 2.0.0 on a RTX 3060 GPU, and follow the training procedure described in Section~\ref{pro-mec}. The uplink transmit power $P_u$ is fixed within each training process, and a separate model is trained for different considered power setting. Specifically, we separately train the models at $20$~dBm and $13~$dBm for the corresponding comparisons, while all other evaluations reuse the models trained at $P_u = 20$~dBm. The DL-based baselines are also trained under the same transmit power setting as the proposed models.

\subsection{Baselines}

To evaluate the effectiveness of the proposed framework, we consider the following representative baselines.

\textit{1) DL-based localization with fixed ERA:}
This baseline uses the same measured finite-state codebook as Model~II, but the ERA state is kept unchanged during the sensing process. Specifically, we choose a fixed state from the measured codebook whose radiation response provides a relatively high gain over the angular support of the considered UE distribution. In our implementation, the same selected state is applied to all antenna elements, pilot blocks, and sensing stages.
Therefore, this baseline preserves the directional radiation property of a realizable ERA state, but does not perform active EM-domain reconfiguration. This baseline is used to distinguish the performance gain brought by adaptive ERA reconfiguration from the gain brought by using a fixed directional ERA pattern.

\begin{figure}[!t]
	\centering
	\includegraphics[width=0.8\linewidth]{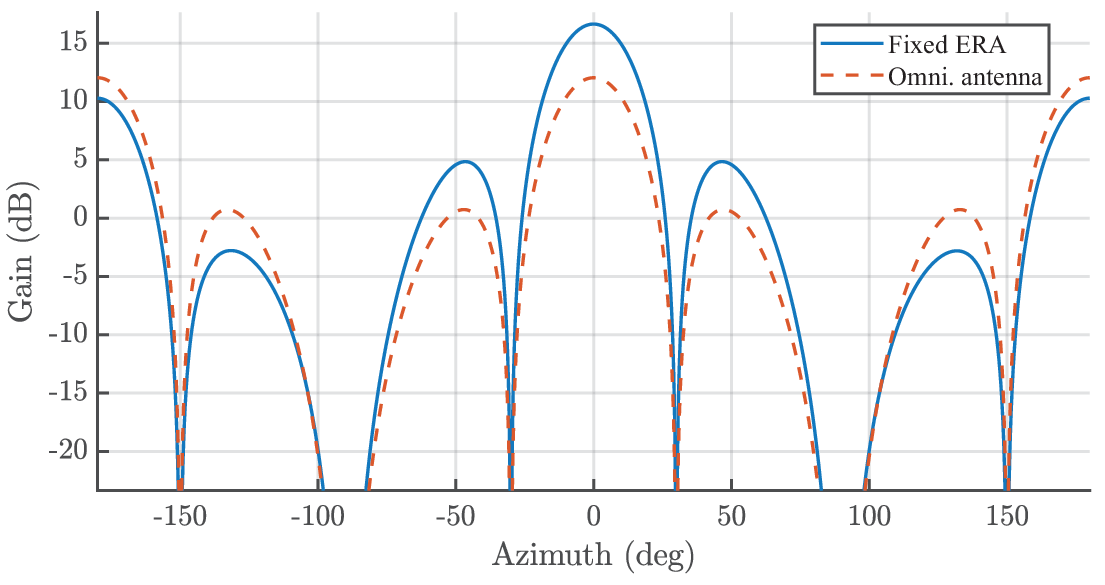}
	\caption{Effective gain comparison between the fixed-ERA and omnidirectional antenna baselines under the setup in Table~\ref{tab:simulation_parameters}.}
	\label{fig:base_pattern}
	\vspace*{-1.0em}
\end{figure}

\textit{2) DL-based localization with omnidirectional antenna:}
This baseline replaces the ERA elements with conventional omnidirectional antenna elements. Therefore, the element-wise EM response is direction-independent and remains fixed over all pilot blocks and sensing stages, which can be written as \( \mathbf{g}^{\rm omni}_{t,p}(\boldsymbol{\vartheta})=\mathbf{1}_{N}\). 
This baseline represents a conventional fixed-pattern antenna array and is used to evaluate the localization gain provided by directional ERA patterns and adaptive EM-domain sensing. Fig.~\ref{fig:base_pattern} compares the azimuth-domain effective gains of the two fixed-pattern baselines (computing by $\vert\mathbf{a}^{\mathsf H}(\boldsymbol{\vartheta})\cdot\mathbf{g}(\boldsymbol{\vartheta})\vert^2$), whereas the fixed ERA provides a relatively stronger directional response over the considered angular region.

\textit{3) DL-based localization with random ERA configurations:}
This baseline uses the same measured finite-state codebook and DL-based localization network as Model~II, but replaces the learnable ERA configuration with random policy. At each stage, the states for all ERAs and pilot blocks are randomly sampled from the measured codebook and resampled for the next stage. Therefore, this baseline provides stage-wise EM-domain measurement diversity but does not perform observation-adaptive sensing. 
This comparison isolates the gain of learned ERA adaptation from that provided by random ERA reconfiguration and the DL-based localization module.

\textit{4) Codebook-based active sensing with ERA:}
In each stage, UE-specific observations are first separated by pilot matching, as in the proposed method. The azimuth/elevation angle of each UE is then estimated using multiple signal classification (MUSIC) method, where the scan grid covers azimuth \([-90^\circ,90^\circ]\) with 181 points and elevation \([-55^\circ,55^\circ]\) with 111 points. The delay/range is then obtained from an oversampled inverse-DFT delay profile with an oversampling factor of 8. Since the UE labels are preserved after pilot matching, the angle and delay estimates are paired within the same pilot label. For next stage, the estimated UE directions are assigned to the \(P\) pilot blocks in a round-robin manner. Each block selects the measured ERA state with the largest interpolated gain toward its assigned direction, and the selected state is applied to all antenna elements\footnote[4]{For each pilot block, a common state is applied to all antenna elements to provide a low-complexity benchmark, since independent per-element selection would introduce a much larger combinatorial design problem.}. If no valid direction estimate is available, the broadside direction is used as the fallback.

\textit{5) Model-based localization with omnidirectional antenna array:}
The BS equipped with conventional omnidirectional antenna collects all pilot observations and then applies a MUSIC-based angular estimator and DFT-based delay estimation to obtain the user angles and ranges for localization.
This baseline serves as a non-adaptive model-based benchmark using an omnidirectional antenna array.

\textit{6) Ablations:} We consider three variants of the proposed framework by removing the GNN module, the LSTM module, and learned attention pooling, respectively. Without the GNN, user representations are processed independently before global aggregation; without the LSTM, the current-stage observation features are directly fed to the GNN without recurrent memory; and without attention pooling, the shared ERA context is obtained by uniformly averaging the active-user embeddings.

To maintain the consistency of the comparison, baselines~\textit{1)} -- \textit{3)} follow the same transmission process as the proposed framework. At each sensing stage, the BS collects pilot-block observations under the fixed/random ERA response, and the observations are processed stage-by-stage by a DL-based localization network. In addition, ablations are also constructed from the Model~II mechanism, where the ERA-related operations follow the discrete state-selection model. For each following evaluation experiment, we generate another $5,120$ independent samples and evaluate the fully trained proposed framework and all baselines using a batch size of $128$. All compared schemes use the same geometric channel distribution, active-user masks, and total pilot-block budget.

\vspace*{-0.85em}
\subsection{Training Convergence}
\vspace*{-0.15em}
\begin{figure}[!t]
	\centering
	\begin{subfigure}[b]{0.24\textwidth}
		\centering
		\includegraphics[width=\textwidth,height=0.69\textwidth]{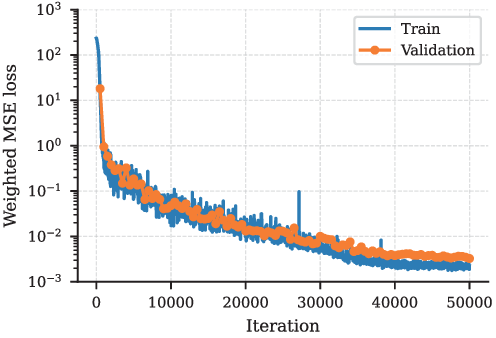}
		\caption{}
		\label{fig:training_model1}
	\end{subfigure}
	\hfill
	\begin{subfigure}[b]{0.24\textwidth}
		\centering
		\includegraphics[width=\textwidth,height=0.69\textwidth]{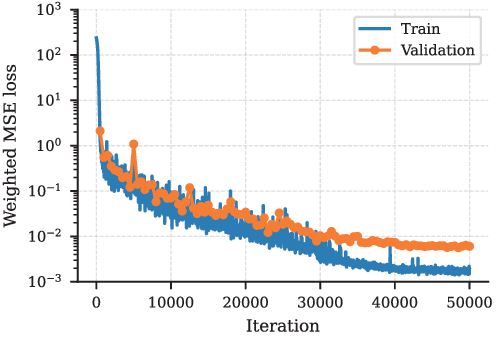}
		\caption{}
		\label{fig:training_model2}
	\end{subfigure}
	\hfill
	\begin{subfigure}[b]{0.24\textwidth}
		\centering
		\includegraphics[width=\textwidth,height=0.69\textwidth]{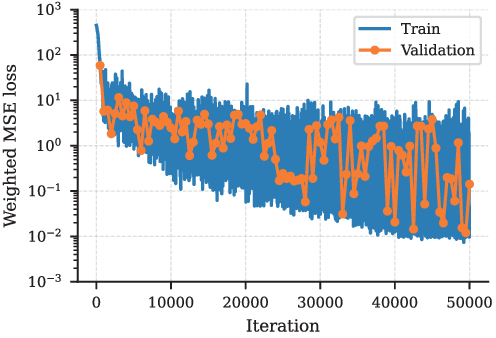}
		\caption{}
		\label{fig:training_ta}
	\end{subfigure}
	\hfill
	\begin{subfigure}[b]{0.24\textwidth}
		\centering
		\includegraphics[width=\textwidth,height=0.69\textwidth]{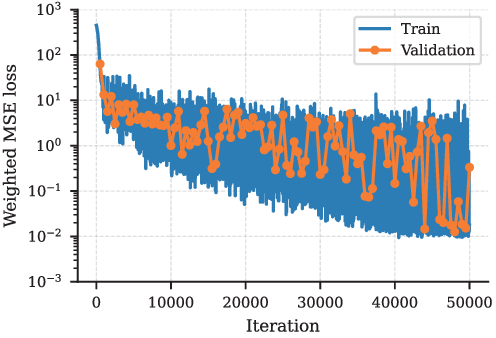}
		\caption{}
		\label{fig:training_oa}
	\end{subfigure}
	\caption{Training convergence comparison under (a) Model I, (b) Model II, (c) Fixed ERA baseline, (d) Omnidirectional antenna baseline.}
	\label{fig:training_comparison}
	\vspace{-1.0em}
\end{figure}

We first show the convergence behavior of the proposed framework and the two fixed-pattern baselines in Fig.~\ref{fig:training_comparison}.
As shown in Fig.~\ref{fig:training_model1}, under the synthesis model, the proposed framework converges rapidly in the early stage and both the training and validation losses gradually approach a low stable level, demonstrating favorable optimization behavior under the idealized ERA configuration methodology. For the measured finite-state model in Fig.~\ref{fig:training_model2}, the overall convergence trend remains largely consistent, although a moderate gap between training and validation appears in the later iteration. This indicates some loss in generalization compared with the synthesis model, which is expected since Model~II is constrained by the measured finite-state radiation codebook and therefore represents a more practical but less flexible hardware configuration than Model~I.
Importantly, the validation loss still decreases and eventually stabilizes, confirming that Model II remains practically effective.
By comparison, the fixed-pattern baselines in Fig.~\ref{fig:training_ta} and Fig.~\ref{fig:training_oa} exhibit substantially larger fluctuations during training. 
Although their losses still follow an overall decreasing trend, their validation losses fluctuate more significantly and do not converge as smoothly as those of the proposed ERA-assisted active sensing models. One possible explanation is that the non-adaptive responses provide less measurement diversity, which may make the localization mapping more difficult to optimize. 
Without adaptive ERA configuration, the network has to infer the positions under fixed spatial responses, making it more sensitive to varying number of UEs, difficult geometries, and multipath conditions. 

\vspace*{-0.8em}
\subsection{Localization Performance and Ablation Comparisons}\label{Per}
\vspace*{-0.2em}

\begin{figure}[t!]
	\centering 
	\includegraphics[width=0.436\textwidth]{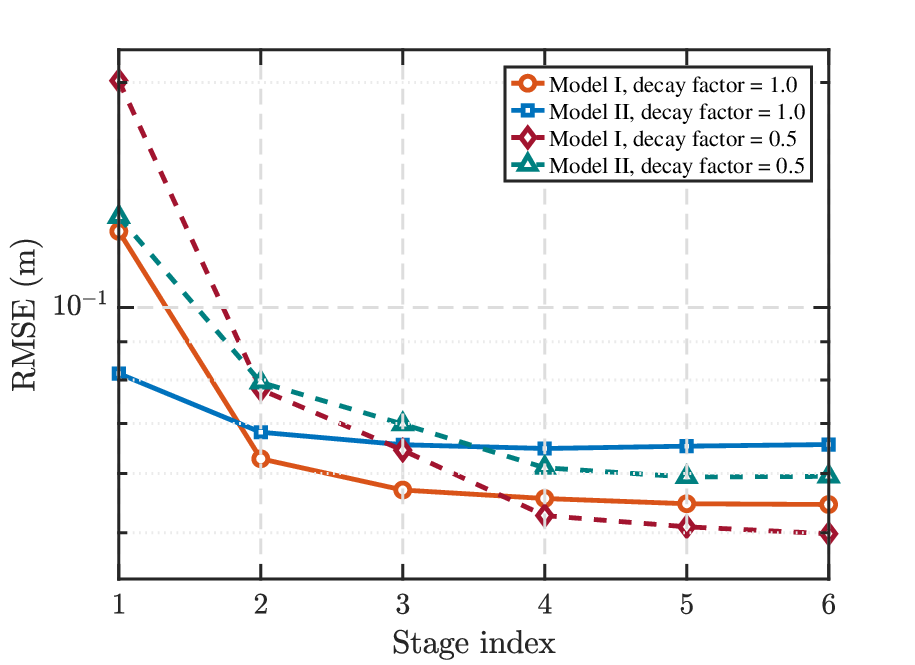} 
	\caption{Localization RMSE versus sensing stage under different stage-weight decay factors.}
	\label{fig:com_decay}
\end{figure}

We first examine whether the network should prioritize uniformly reliable position estimates across all stages or mainly optimize the final-stage estimate.
To provide an intuitive measure of the localization error, we adopt the root mean square error (RMSE) as the performance metric. The stage-wise weights $\{\alpha_t\}_{t=1}^{T}$ is controlled by a decay factor $\gamma$, where $\alpha_t = \frac{\gamma^{T-t}} {\sum_{i=1}^{T}\gamma^{T-i}}$. 
When $\gamma=1$, all stages are equally weighted;
when $0<\gamma<1$, larger weights are assigned to later stages.
As shown in Fig.~\ref{fig:com_decay}, 
compared with the equal-weight case, using a smaller decay factor can slightly improve the accuracy at the final stage, but it also leads to degraded early stage performance. 
Under limited initial observations, high EM-domain flexibility allows the network to allocate the early stage toward broad spatial exploration rather than immediate fine localization. 
After more observations are accumulated, however, the EM-domain flexibility enables to refine the localization more effectively, leading to improved RMSE in the later stages. These results confirm the expected tradeoff introduced by the stage-wise objective in~\eqref{prob}. Since the proposed framework is intended to provide progressively refined estimates and may be terminated before the maximum number of stages, we use $\gamma=1$ in all subsequent experiments. 

\begin{figure}[!t]
	\centering
	\begin{subfigure}[b]{0.406\textwidth}
		\centering
		\includegraphics[width=\textwidth,height=0.73\textwidth]{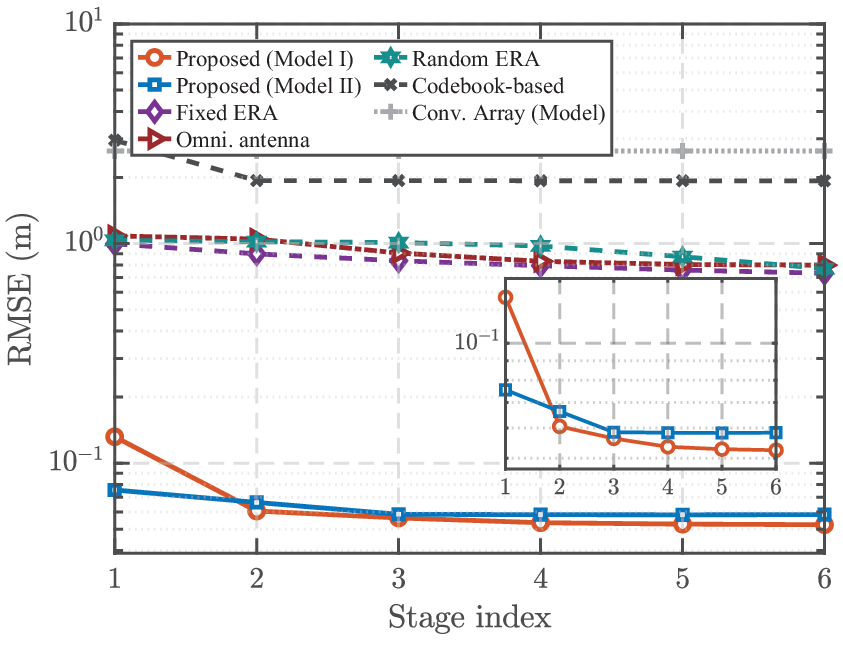}
		\caption{Localization performance under $P_u=$ 20dBm.}
		\label{fig:stage1}
	\end{subfigure}
	\hfill
	\begin{subfigure}[b]{0.411\textwidth}
		\centering
		\includegraphics[width=\textwidth,height=0.73\textwidth]{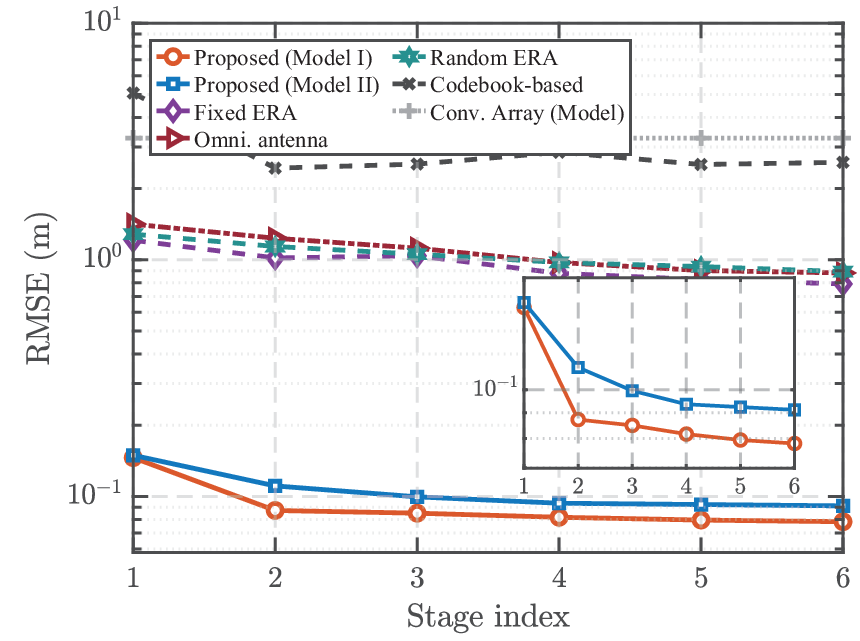}
		\caption{Localization performance under $P_u=$ 13dBm.}
		\label{fig:stage2}
	\end{subfigure}
	\caption{Localization RMSE versus sensing stage under different uplink transmit power $P_u$.}
	\label{fig:comparison_S}
	\vspace{-1.0em}
\end{figure}

Fig.~\ref{fig:comparison_S} shows the localization RMSE versus the sensing stage under different uplink transmit powers. It can be observed that the ERA-assisted active sensing framework achieves a clear and progressive RMSE reduction as the stage increases. This improvement is especially significant in the first few stages, which indicates that the proposed active sensing framework can effectively exploit the accumulated pilot observations to refine the subsequent sensing strategy. It should be noted that the first-stage gain is attributed to the predefined structured probing and the supervised localization network, rather than ERA configuration. Adaptive refinement starts from the second stage and is reflected in the subsequent RMSE reduction and the advantage over the fixed and random ERA baselines.

Moreover, Model I consistently achieves the best performance, while Model II remains close to Model I over all stages. This is reasonable since Model I relies on the synthesis-based ERA model with continuous EM-domain control, whereas Model~II is constrained by the measured finite-state radiation codebook. Nevertheless, Model~II still outperforms the baselines, demonstrating its effectiveness as a practically deployable implementation. 
When the transmit power decreases from $20$ dBm to $13$ dBm, the RMSE of all methods increases, but the proposed Model I and Model II still preserve clear performance advantages. 
The results of the fixed ERA, random ERA, and omnidirectional antenna baselines clarify the role of EM-domain pattern design and observation-adaptive reconfiguration. The fixed-ERA baseline outperforms the omnidirectional baseline due to its selected directional gain, while the random ERA baseline remains close to the fixed-ERA baseline despite providing stage-wise pattern diversity. This indicates that random reconfiguration alone is insufficient, and that the main gain of Model~II comes from adapting ERA configuration.
The gradual RMSE reduction of the non-adaptive baselines is mainly attributed to LSTM-based observation accumulation rather than active ERA adaptation.

\begin{figure}[t!]
	\centering 
	\includegraphics[width=0.436\textwidth]{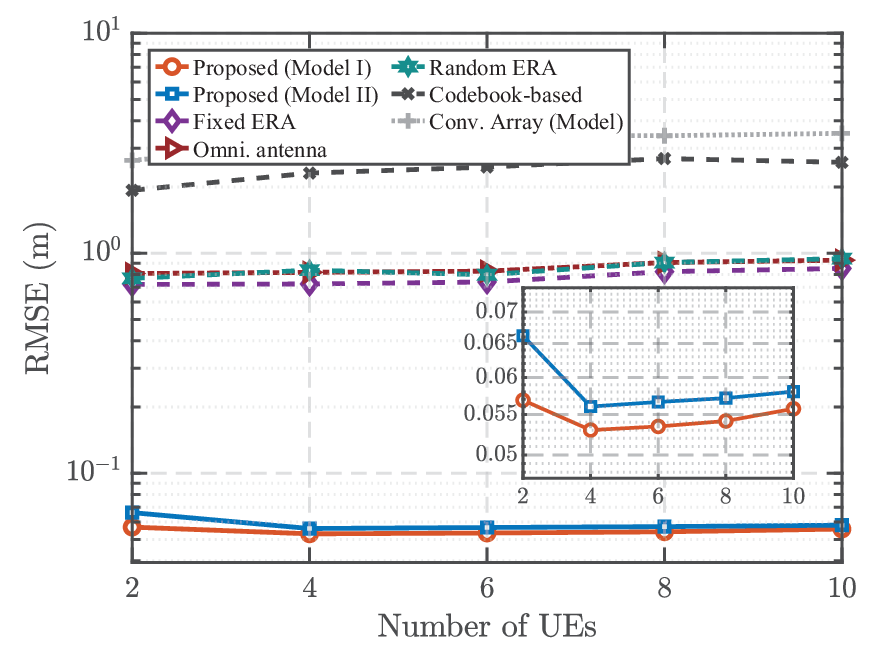} 
	\caption{Localization RMSE versus number of UEs.}
	\vspace{-1.0em}
	\label{fig:com_K}
\end{figure}


Fig.~\ref{fig:com_K} shows the localization RMSE versus the number of UEs. The proposed Model~I and Model~II maintain nearly stable RMSE as the number of UEs increases, indicating that the learned framework can accommodate varying user counts within the considered range. Model~I consistently achieves the lowest RMSE, while Model~II remains comparable, showing that the measured finite-state ERA preserves most of the localization gain provided by the ERAs. In contrast, the fixed ERA, random ERA, and omnidirectional antenna baselines exhibit substantially larger RMSEs. 
The codebook-based and model-based baselines perform even worse, highlighting the limitations of hand-crafted angle/delay estimation. 
Overall, the results demonstrate that the proposed framework generalizes well to different numbers of UEs within the considered channel distribution and user count range.


\begin{figure}[t!]
	\centering 
	\includegraphics[width=0.476\textwidth]{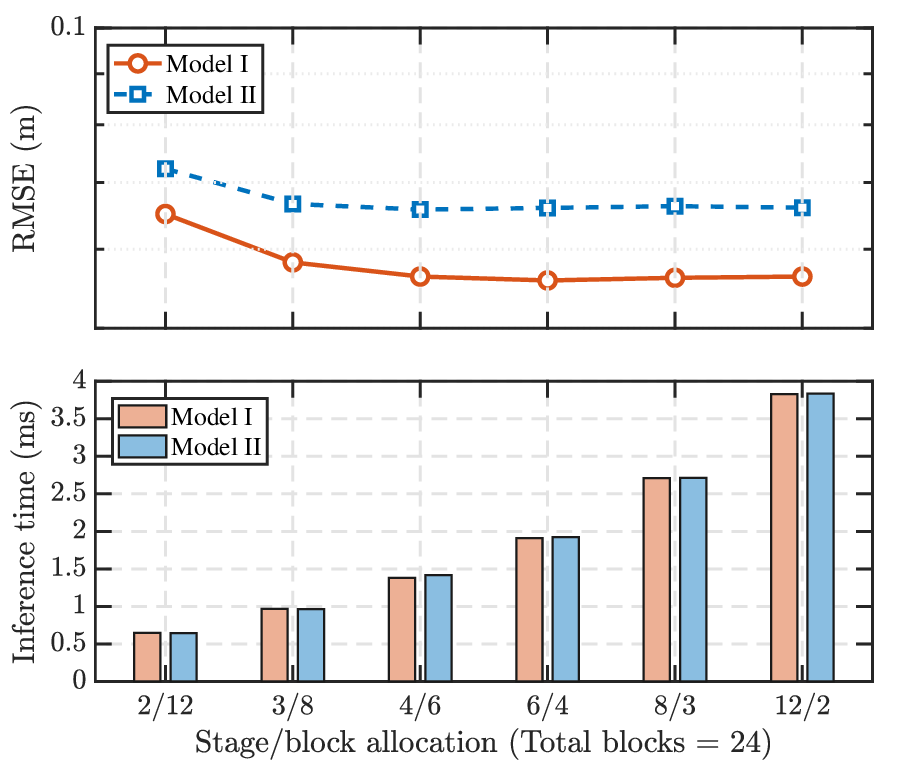} 
	\caption{Localization RMSE and per-sample neural inference time versus stage/block allocation.}
	\vspace{-1.0em}
	\label{fig:com_alloc}
\end{figure}

Fig.~\ref{fig:com_alloc} compares the localization RMSE and per-sample neural inference time under different stage/block allocations, with the total pilot-block budget fixed at $24$. We train and evaluate separate models under each allocation, and per-sample inference time is obtained by averaging the forward-pass latency over all samples in one batch. As shown in the upper figure, increasing the number of sensing stages generally improves the RMSE for both models, since the sensing strategy can be updated frequently. The improvement is most significant from the $2/12$ allocation to the $4/6$ allocation, while the RMSE curves become stable when the number of stages further increases.
The lower figure shows that the per-sample inference time increases with the number of stages, because more stages require more neural processing. Nevertheless, both Model~I and Model~II achieve millisecond-level neural inference latency under all considered allocations. Although the reported time only measures inference latency for one sample on the adopted computing platform and does not include signal transmission or hardware-control delays, the millisecond-level inference results still indicate that the proposed framework has relatively low computational overhead.
Overall, Fig.~\ref{fig:com_alloc} reveals a tradeoff between localization accuracy and computational complexity. Under the considered setting and channel distribution, the $4/6$ allocation provides a favorable balance between performance and inference cost.

To evaluate the robustness of the proposed framework to practical mismatches, we further test the trained Model~II under LoS blockage, measured-pattern perturbations, and synchronization errors. For the LoS blockage, the LoS path of each UE is independently retained with probability $p_{\mathrm{LoS}}$. For the pattern mismatch, an independent Gaussian perturbation is applied to each angular gain sample of each measured finite-state pattern. Specifically, the perturbed amplitude response is generated as $\widetilde{b}_{s}(\boldsymbol{\vartheta})=b_{s}(\boldsymbol{\vartheta}) 10^{\epsilon_{s,\boldsymbol{\vartheta}}/20}$, where $\epsilon_{s,\boldsymbol{\vartheta}} \sim \mathcal{N}(0,\sigma_{\mathrm{pat}}^{2})$ denotes the independent zero-mean Gaussian perturbation, where $\sigma_{\mathrm{pat}}$ controls the perturbation strength. For the synchronization error, each UE is assigned an uncompensated clock bias $\Delta\tau_u\sim\mathcal{N}(0,\sigma_{\mathrm{clk}}^2)$ during testing.

\begin{figure}[t!]
	\centering 
	\includegraphics[width=0.426\textwidth]{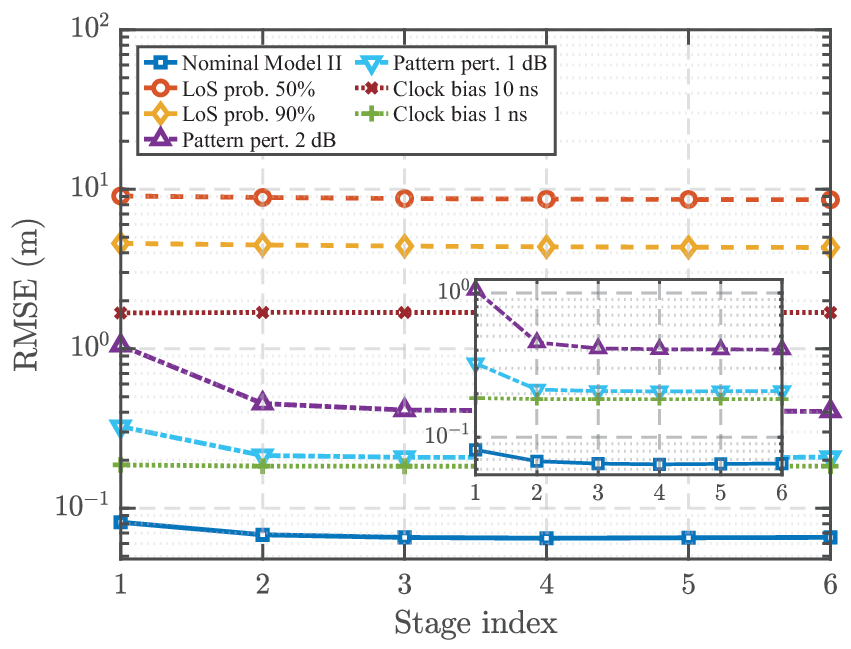} 
	\caption{Localization RMSE versus sensing stage under practical mismatches.}
	\vspace{-1.0em}
	\label{fig:model_mismatch}
\end{figure}


Fig.~\ref{fig:model_mismatch} evaluates the stage-wise localization performance of Model~II under several mismatches. Measurement perturbations increase the RMSE as the perturbation strength grows, but both the $1$-dB and $2$-dB cases still exhibit clear improvement over the first few sensing stages. This indicates that the learned framework can continue to exploit multi-stage observations under moderate codebook mismatch. In addition, a $1$-ns clock bias causes a moderate performance degradation, whereas a $10$-ns bias leads to a larger localization error. In both cases, the RMSE remains nearly unchanged across stages, indicating that multi-stage observation accumulation cannot compensate for a systematic delay. LoS blockage leads to meter-level errors with limited stage-wise refinement, because removing the LoS component significantly changes the learned angle-delay relationship. Overall, the proposed framework exhibits partial robustness to small clock bias and moderate pattern perturbations, but remains sensitive to large synchronization errors and LoS blockage, motivating further mismatch modeling or training augmentation.

\begin{figure}[t!]
	\centering 
	\includegraphics[width=0.406\textwidth]{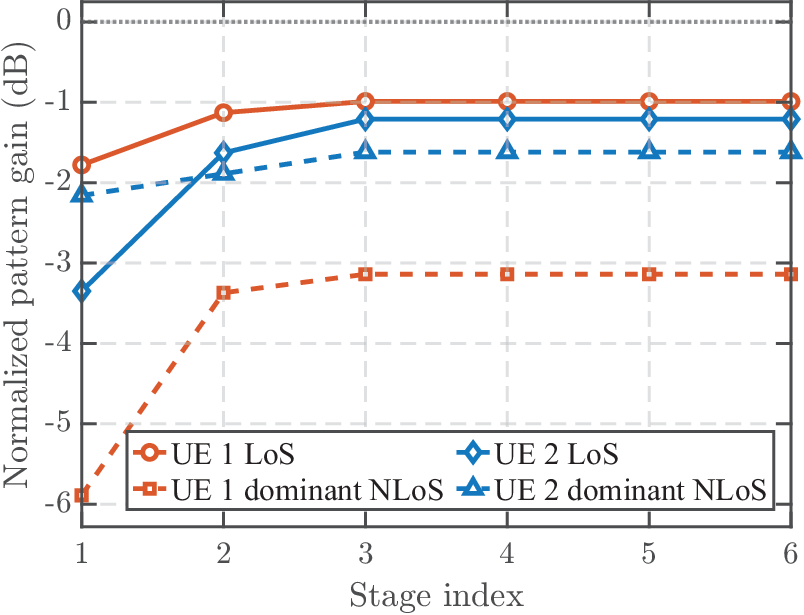} 
	\caption{Average normalized pattern gains toward different directions over sensing stages in a two-UE scenario.}
	\vspace{-1.0em}
	\label{fig:pattern_tracking}
\end{figure}

To provide a diagnostic view of the learned ERA configuration, we examine the normalized effective pattern gain toward the LoS and dominant NLoS directions under a two-UE scenario with 5,120 samples. For each block, the effective pattern is normalized by its maximum angular power response, and the resulting gains are averaged over all test samples and pilot blocks. The dominant NLoS path is defined as the NLoS component with the largest path amplitude for each UE.
As shown in Fig.~\ref{fig:pattern_tracking}, the normalized gains toward both the LoS and dominant NLoS directions generally increase during the first few sensing stages and then become stable. In particular, the learned policy maintains enhanced relative responses toward multiple propagation directions rather than concentrating exclusively on a single UE or path.
It is worth noting that the measured finite-state ERA patterns are predominantly broadside-oriented, with their main lobes generally formed around $0^\circ$~\cite{era-loc1}, as also illustrated in Fig.~\ref{fig:base_pattern}. Consequently, the selected ERA states cannot in general place their maximum response exactly along arbitrary LoS or NLoS direction. Therefore, the path gains in Fig.~\ref{fig:pattern_tracking} remain below $0$ dB even after adaptation. Their progressive increase should be interpreted as stronger relative emphasis on these informative propagation directions. Overall, the results indicate that the adaptive ERA policy progressively reallocates the shared-aperture response toward LoS and multipath components.

\begin{table}[t]
	\centering
	\caption{Final-stage RMSE degradation of ablations.}
	\label{tab:ablation_final_stage}
	\begin{tabular}{lcc}
		\toprule
		Variant & Degradation at $20$ dBm & Degradation at $13$ dBm \\
		\midrule
		w/o GNN   & 4.3\%  & 10.7\% \\
		w/o LSTM  & 13.1\% & 36.4\% \\
		w/o Attn. & 6.1\%  & 16.8\% \\
		\bottomrule
	\end{tabular}
	\vspace*{-1.0em}
\end{table}

Table~\ref{tab:ablation_final_stage} summarizes the localization RMSE degradation of the architectural variants relative to the full Model~II. Removing the LSTM leads to the largest degradation, particularly at $13$ dBm, indicating that temporal accumulation of multi-stage observations becomes more important under lower-SNR conditions. Replacing attention pooling with mean pooling causes the second-largest performance loss, while removing the GNN results in a smaller but consistent degradation. Overall, the ablation results show that all three modules contribute to the proposed framework, with the recurrent memory providing the most pronounced benefit. These architectural gains, however, remain secondary to the larger system-level improvement brought by observation-adaptive ERA reconfiguration.

\section{Conclusion}\label{col}

In this paper, we investigated ERA-assisted multi-user localization in a wideband uplink system via learning-based active sensing. A unified framework was developed to accommodate both the synthesis-based and the finite-state model. To solve the resulting coupled problem, we proposed a deep active sensing architecture integrating LSTM-based state accumulation, GNN-based multi-user interaction, and adaptive weighting. Numerical results showed that the proposed method outperforms representative benchmarks and achieves refinement across sensing stages. Moreover, the measured-based modeling paradigm obtained performance close to the synthesis-based counterpart, demonstrating its potential practical applicability. 

However, although the present results are promising, they should be interpreted within the considered channel distribution and under ideal training assumptions. Specificity, the main framework assumes synchronized clocks and calibrated finite-state radiation patterns. Moreover, the reported latency accounts only for neural inference and excludes pilot transmission and hardware delays. Although the robustness experiments provide an initial evaluation under LoS blockage, clock bias, and pattern perturbations, the model is still designed under nominal synchronized and calibrated conditions. Extending the framework to mismatch-aware training and experimental hardware validation constitutes important future work.


\begin{thebibliography}{1}
\bibliographystyle{IEEEtran}
	\bibitem{b1}
	S. E. Trevlakis et al., “Localization as a Key Enabler of 6G Wireless Systems: A Comprehensive Survey and an Outlook,” \textit{IEEE Open J. Commun. Soc.,} vol. 4, pp. 2733-2801, 2023.
	
		
	\bibitem{b2}
	X. Cai, X. Cheng and F. Tufvesson, “Toward 6G with Terahertz Communications: Understanding the Propagation Channels,” \textit{IEEE Commun. Mag.,} vol. 62, no. 2, pp. 32-38, Feb. 2024.
	
	\bibitem{b3}
	K. Witrisal et al., “High-accuracy localization for assisted living: 5G systems will turn multipath channels from foe to friend,” \textit{IEEE Signal Process. Mag.,} vol. 33, no. 2, pp. 59-70, Mar. 2016.
	
	\bibitem{b4}
	X. Mu, Y. Liu, L. Guo, J. Lin, and R. Schober, “Intelligent reflecting surface enhanced indoor robot path planning: A radio map-based approach,” \textit{IEEE Trans. Wireless Commun.,} vol. 20, no. 7, pp. 4732-4747, Jul. 2021.
	
	
	\bibitem{b6} S. Aditya, A. F. Molisch and H. M. Behairy, "A Survey on the Impact of Multipath on Wideband Time-of-Arrival-Based Localization," \textit{Proc. IEEE,} vol. 106, no. 7, pp. 1183-1203, Jul. 2018.
	
	\bibitem{m1} F. Rusek et al., “Scaling up MIMO: Opportunities and challenges with very large arrays,” \textit{IEEE Signal Process. Mag.,} vol. 30, no. 1, pp. 40-60, Jan. 2013.
		
	\bibitem{m2} Q. Xue et al., “A	Survey of Beam Management for mmWave and THz Communications Towards 6G,” \textit{IEEE Commun. Surveys Tuts.,} vol. 26, no. 3, pp. 1520-1559, 2024.
	
	
	\bibitem{m4} K. Ying et al., “Reconfigurable Massive MIMO: Precoding Design and Channel Estimation in the Electromagnetic Domain,” \textit{IEEE Trans. Commun.,} vol. 73, no. 5, pp. 3423-3440, May 2025.
			
	\bibitem{m5}
	B. Zhou, A. Liu and V. Lau, “Successive Localization and Beamforming in 5G mmWave MIMO Communication Systems,” \textit{IEEE Trans. Signal Process.,} vol. 67, no. 6, pp. 1620-1635, 15 March, 2019.
	
	\bibitem{era1} R. Wang et al., “Electromagnetically Reconfigurable Fluid Antenna System for Wireless Communications: Design, Modeling, Algorithm, Fabrication, and Experiment,” \textit{IEEE J. Sel. Areas Commun.,} vol. 44, pp. 1464-1479, 2026.
	
	\bibitem{era4} J. Zhang et al., “A novel pixel-based reconfigurable antenna applied in fluid antenna systems with high switching speed,” \textit{IEEE Open J. Antennas Propag.,} vol. 6, no. 1, pp. 212-228, Feb. 2025.
	
	\bibitem{era5} M. Liu et al., “Tri-timescale beamforming design for tri-hybrid architectures with reconfigurable antennas,” \textit{arXiv preprint arXiv:2503.03620,} 2025.
	
	\bibitem{era6} Z. Han et al., “Characteristic mode analysis of ESPAR for single-RF MIMO systems,” \textit{IEEE Trans. Wireless Commun.,} vol. 20, no. 4, pp. 2353-2367, Apr. 2021.
	
	\bibitem{bc-era1} P. Zheng et al., “Tri-Hybrid Multi-User Precoding Using Pattern-Reconfigurable Antennas: Fundamental Models and Practical Algorithms,” \textit{arXiv preprint arXiv:2505.08938,} 2025.
	
	\bibitem{bc-era2} W. Ma et al., "A Survey on Reconfigurable and Movable Antennas for Wireless Communications and Sensing," \textit{IEEE Commun. Surveys Tuts.,} vol. 28, pp. 4842-4882, 2026.
	
		
	\bibitem{era2} P. Zheng et al., “Electromagnetically Reconfigurable Antennas for 6G: Enabling Technologies, Prototype Studies, and Research Outlook,” \textit{arXiv preprint arXiv:2506.00657,} 2025.
	
	\bibitem{era-loc1} A. Fadakar et al., "Hybrid Codebook Design for Localization Using Electromagnetically Reconfigurable Fluid Antenna System," \textit{IEEE J. Sel. Topics Signal Process.,} early access, pp. 1-16, 2026.
	
	\bibitem{ma1} L. Zhu et al., “A Tutorial on Movable Antennas for Wireless Networks,” \textit{IEEE Commun. Surveys Tuts.,} vol. 28, pp. 3002-3054, 2026.
	
	\bibitem{bc-mfa1} Y. Zhang et al., “6D Movable Antenna-Aided Hybrid Beamforming for Multi-User Communications,” 2024 IEEE Globecom Workshops (GC Wkshps), Cape Town, South Africa, 2024, pp. 1-6.
	
	\bibitem{fa1} W. K. New et al., “A Tutorial on Fluid Antenna System for 6G Networks: Encompassing Communication Theory, Optimization Methods and Hardware Designs,” \textit{IEEE Commun. Surveys Tuts.,} vol. 27, no. 4, pp. 2325-2377, Aug. 2025.
	
	\bibitem{era-aoa1} E. Taillefer et al., “Direction-of-arrival estimation using radiation power pattern with an ESPAR antenna,” \textit{IEEE Trans. Antennas Propag.,} vol. 53, no. 2, pp. 678-684, 2005.
	
	\bibitem{era-aoa2} R. Qian et al., “Direction-of-arrival estimation with single-RF ESPAR antennas via sparse signal reconstruction,” in \textit{2015 IEEE 16th International Workshop on Signal Processing Advances in Wireless Communications (SPAWC),} 2015, pp. 485-489.
	
	\bibitem{era-aoa3} L. Kulas, “Simple 2-D direction-of-arrival estimation using an ESPAR antenna,” \textit{IEEE Antennas Wireless Propag. Lett.} vol. 16, pp. 2513-2516, 2017.
	
	\bibitem{era-loc2} M. Rzymowski et al., “Single-anchor indoor localization using ESPAR antenna,” \textit{IEEE Antennas Wireless Propag. Lett.,} vol. 15, pp. 1183-1186, 2016.
	
	
	
	\bibitem{az2} S.-E. Chiu et al., “Active learning and CSI acquisition for mmWave initial alignment,” \textit{IEEE J. Sel. Areas Commun.,} vol. 37, no. 11, pp. 2474-2489, Nov. 2019.
	
	\bibitem{az3} F. Sohrabi et al., “Active sensing for communications by learning,” \textit{IEEE J. Sel. Areas Commun.,} vol. 40, no. 6, pp. 1780-1794, Jun. 2022.
	
	\bibitem{az4} T. Jiang, F. Sohrabi, and W. Yu, “Active sensing for two-sided beam alignment using ping-pong pilots,” in \textit{Proc. Asilomar Conf. Signals Syst. Comput.,} Pacific Grove, California, USA, Nov. 2022, pp. 913-918.
	
	\bibitem{az5} T. Jiang, and W. Yu, “Active sensing for reciprocal MIMO channels,” \textit{IEEE Trans. Signal Process.,} vol. 72, pp. 2905-2920, 2024.
	
	\bibitem{a1} R. Zhang, Y. Zhang, and Y. Zhang, “User Localization via Active Sensing with Electromagnetically Reconfigurable Antennas,” \textit{arXiv preprint arXiv:2601.20501,} 2026.
	
	\bibitem{ar1} Y. Li and W. Yu, “Localization in multipath environments via active sensing with reconfigurable intelligent surfaces,” \textit{IEEE Commun. Lett.,} vol. 28, no. 9, pp. 2061-2065, Sep. 2024.
	
	\bibitem{ar2} Z. Zhang, T. Jiang, and W. Yu, “Active sensing for localization with reconfigurable intelligent surface,” in \textit{Proc. IEEE Int. Conf. Commun. (ICC),} Rome, Italy, May 2023, pp. 4261-4266.
	
	\bibitem{ar3} Z. Zhang, T. Jiang, and W. Yu, “Localization with reconfigurable intelligent surface: An active sensing approach,” \textit{IEEE Trans. Wireless Commun.,} vol. 23, no. 7, pp. 7698-7711, Jul. 2024.
	
	\bibitem{ar6} Z. Zhang and W. Yu, "Learning Beamforming Codebooks for Active Sensing With Reconfigurable Intelligent Surface," \textit{IEEE Trans. Wireless Commun.,} vol. 24, no. 8, pp. 6504-6517, Aug. 2025.
	
	\bibitem{a6} H. Han, T. Jiang, and W. Yu, “Active Sensing for Multiuser Beam Tracking With Reconfigurable Intelligent Surface,” \textit{IEEE Trans. Wireless Commun.,} vol. 24, no. 1, pp. 540-554, Jan. 2025.
		
	\bibitem{gnn1} Y. Shen, et al., “Graph neural networks for scalable radio resource management: Architecture design and theoretical analysis,” \textit{IEEE J. Sel. Areas Commun.,} vol. 39, no. 1, pp. 101--115, Jan. 2021.
		
			
	\bibitem{bc1} J. Chen et al., “Integrated Sensing and Communication with Tri-Hybrid Beamforming Across Electromagnetically Reconfigurable Antennas,” \textit{arXiv preprint, arXiv:2510.14530,} 2025.
	
	\bibitem{OD} M. Costa et al., “Unified array manifold decomposition based on spherical harmonics and 2-D fourier basis,” \textit{IEEE Trans. Signal Process.,} vol. 58, no. 9, pp. 4634-4645, 2010.
	
	\bibitem{bc2} S. Liang and R. Srikant, “Why deep neural networks for function
	approximation?” in \textit{Proc. Int. Conf. Learn. Represent. (ICLR),} Toulon,
	France, 2017.
	
	\bibitem{bc3} W. Yu et al., “Role of deep learning in wireless communications,” \textit{IEEE BITS Inf. Theory Mag.,} vol. 2, no. 2, pp. 56-72, Nov. 2022.
	
	\bibitem{deepsets} M. Zaheer, et al., “Deep Sets,” 	\textit{Adv. Neural Inf. Process. Syst. (NeurIPS),} 	vol. 30, pp. 3391--3401, 2017.
		
	\bibitem{settransformer} J. Lee, et al., “Set Transformer: A framework for attention-based permutation-invariant neural networks,” \textit{Proc. 36th Int. Conf. Mach. Learn. (ICML),} vol. 97, pp. 3744--3753, Jun. 2019.
	
%
	\bibitem{bc5} T. Jiang, H. V. Cheng, and W. Yu, “Learning to reflect and to beamform for intelligent reflecting surface with implicit channel estimation,” \textit{IEEE J. Sel. Areas Commun.,} vol. 39, no. 7, pp. 1931-1945, Jul. 2021.
	
	\bibitem{ste} R. Zhang et al., "A Deep Learning Framework for Joint Channel Acquisition and Communication Optimization in Movable Antenna Systems," \textit{IEEE Trans. Wireless Commun.,} vol. 25, pp. 14471-14485, 2026.
	
	\bibitem{gs} H. Xuan, B. Yang, and X. Li, “Exploring the impact of temperature
	scaling in softmax for classification and adversarial robustness,” \textit{arXiv
	preprint arXiv:2502.20604,} 2025.

	\bibitem{admm} D. P. Kingma and J. Ba, “Adam: A method for stochastic optimization,” \textit{Proc. Int. Conf. Learn. Represent. (ICLR),} San Diego, CA, USA, 2015.
	
	\bibitem{dl} A. Sayal et al., “Neural networks and machine learning,” \textit{Proc. 2023 IEEE 5th Int. Conf. Cybernetics, Cognition Machine Learn. Appl. (ICCCMLA),} Hamburg, Germany, pp. 58-63, 2023.
	
\end{thebibliography}
\end{document}